\documentclass[journal]{IEEEtran}
\ifCLASSINFOpdf
\else
\fi

\usepackage{cite}
\usepackage{amsmath}
\usepackage{amsfonts}
\usepackage{graphicx}
\usepackage{ amssymb }
\usepackage{color}
\usepackage{bbold}
    \usepackage{nicematrix}
    \usepackage{enumitem}

\newcommand{\R}{\mathbb{R}}
\newcommand{\cR}{\mathcal{R}}

\newcommand{\Z}{\mathbb{Z}}
\newcommand{\F}{\mathcal{F}}
\newcommand{\G}{\mathcal{G}}
\newcommand{\oRs}{\overline{\mathcal{R}}_s}
\newcommand{\oR}{\overline{\mathcal{R}}}
\newcommand{\h}{\mathcal{H}}

\newcommand{\sm}{\text{-}} 
\newcommand{\cG}{{\cal G}}
\newcommand{\be}{\begin{equation}}
\newcommand{\ee}{\end{equation}}
\newcommand{\bbm}{\begin{bmatrix}} 
\newcommand{\ebm}{\end{bmatrix}} 
\newcommand{\lb}{\left(} 
\newcommand{\rb}{\right)} 
\newcommand{\first}{1^{\text{st}}}

\newcommand{\lba}{\left[\begin{array}}
\newcommand{\ear}{\end{array} \right]}

\newtheorem{thm}{Theorem}[section]
\newtheorem{lem}[thm]{Lemma}
\newtheorem{prop}[thm]{Proposition}

\newtheorem{defn}{Definition}[section]

\newtheorem{exmp}{Example}[section]
\newtheorem{rem}{Remark}

\begin{document}

\title{An Explicit Parametrization of Closed Loops\\ for Spatially Distributed Controllers\\ with Sparsity Constraints
}

\author{Emily Jensen$^{1}$ and Bassam Bamieh$^{2}$% <-this % stops a space
\thanks{This work is partially supported by NSF award ECCS-1932777 and CMI-1763064.}% <-this % stops a space
\thanks{$^{1}$Emily Jensen is with the department of Electrical and Computer Engineering,
      University of California, Santa Barbara
        {\tt\small emilyjensen@ucsb.edu}}%
\thanks{$^{2}$Bassam Bamieh is with the Department of Mechanical Engineering, University of California, Santa Barbara
        {\tt\small bamieh@ucsb.edu}}%
}

\maketitle

\begin{abstract}
We study the linear time-invariant state-feedback controller design problem for distributed systems. We follow the recently developed System Level Synthesis (SLS) approach and impose locality structure on the resulting closed-loop mappings; the corresponding controller implementation inherits this prescribed structure. In contrast to existing SLS results, we derive an \emph{explicit} (rather than implicit) parameterization of all achievable stabilized closed-loops. This admits more efficient IIR representations of the temporal part of the closed-loop dynamics, and it allows for the $\h_2$ design problem with closed-loop spatial sparsity constraints to be converted to a standard model matching problem, with the number of transfer function parameters scaling linearly with the closed-loop spatial extent constraint. 
 We illustrate our results with two applications: consensus of first-order subsystems and the vehicular platoons problem. In the case of first-order consensus, we provide analytic solutions and further analyze the architecture of the resulting controller implementation. Results for infinite extent spatially-invariant systems are presented to provide insight to the case of a large but finite number of subsystems.
\end{abstract}


% \fi
\IEEEpeerreviewmaketitle

\section{Introduction}

We consider the linear time-invariant (LTI) state-feedback controller design problem for spatially-distributed systems. The controller to be designed is also a distributed system; each subcontroller component has access to only a subset of \emph{local} system information that is shared across the network according to an underlying communication graph. 
The design problem of interest is to synthesize optimal controllers (with respect to some performance measure) subject to structural constraints that account for this limited information sharing architecture. 

In many cases, imposing structural constraints directly on the controller transfer matrix leads to non-convex design problems.  On the other hand,  imposing sparsity constraints on closed-loop mappings  leads to convex optimization problems. Indeed it has long been known that for any LTI system, the set of all achievable and stabilized closed-loop mappings from disturbance to performance output is convex \cite{boyd1991linear}. However, characterization of this convex set requires knowledge of the open-loop zeros and their directions, leading to complex computation requirements. 

The recently developed System Level Synthesis (SLS) methodology provides a new way of looking at the closed-loop design problem by parameterizing instead a different closed-loop mapping: from specific filtered disturbances to state and control action. In the state-feedback setting, SLS provides an implicit affine subspace constraint to parameterize the set of these alternate achievable stabilized closed-loop mappings using only a state-space realization of the plant. Additional structural constraints on these closed loops (such as sparsity structure) preserve convexity of the design set.  The SLS methodology also proposes controller implementations that inherit the prescribed structure of the closed-loop mappings, and is thus an indirect method for structured controller design.

A potential limitation of the existing SLS literature is that the (infinite-dimensional) \emph{implicit} affine subspace constraint utilized is typically enforced numerically, requiring temporal FIR approximations. Negative effects of these approximations have been shown in \cite{zheng2020}, e.g. infeasibility of affine subspace constraints when dealing with stable uncontrollable, unobservable modes \cite[Sec. 5]{zheng2020}.
The main contribution of this paper is the derivation of an \emph{explicit} (rather than implicit) parameterization of all achievable stabilized closed-loop mappings to allow for analytic IIR solutions. 

Alternate explicit parameterizations of achievable stabilized closed-loops are well-known. For instance, the standard Youla parameterization provides a convex reformulation of the controller design problem subject to structural constraints in special cases (e.g. quadratic invariance \cite{rotkowitz2005characterization} and funnel causality \cite{bamieh2005convex}); however this problem remains non-convex in general settings. In addition, the Youla parameterization requires an explicit computation of a co-prime factorization of the plant, while our approach appears to have such factorizations built into the procedure. This argument is outlined in Section~\ref{sec:youla_connection} for certain special cases.

We focus on the setting of distributed systems for which the dynamics of the individual subsystems are \emph{decoupled} in open loop. This architecture is common in practice; relevant applications include the consensus of first-order subsystems or the vehicular platoons problem \cite{jovanovic2005ill}, \cite{pates2017control}, and vehicular formation problems in general; coupling of the dynamics is introduced through feedback and is not inherent in the open-loop system.
As illustrative examples, we apply our parameterizations to derive analytic solutions to the closed-loop structured $\h_2$ controller design problem for first-order consensus; numerical results are illustrated for the vehicular platoon problem.

 We study the infinite spatial extent spatially-invariant setting to provide insight to the large but finite setting, as in e.g. \cite{bamieh2002distributed, jovanovic2005ill}. Our approach allows us to provide commentary about the performance limitations and scalings with system size for these applications, complementing analysis of recent works including \cite{bamieh2012coherence, tegling2019fundamental, oral2019disorder}.

Our parameterization allows for the optimal $\h_2$ controller design problem with closed-loop sparsity constraints to be reformulated as a standard model-matching problem \cite{francis1987course}. This allows for efficient numerical solutions, and in certain cases analytic solutions may be derived. The number of transfer function parameters in this model matching problem scales linearly with the spatial sparsity constraint on the closed-loop and is finite even in the infinite spatial-extent spatially-invariant setting (provided finite sparsity constraints are imposed).

The rest of the paper is structured as follows. In Section~\ref{sec:notation} we provide preliminaries on spatially-invariant systems and introduce a notion of system locality. In Section~\ref{sec:setup} we formulate the closed-loop structured optimal controller design problem. Our main result is the derivation of an explicit parameterization of all achievable stabilized closed-loop mappings. This parameterization is derived for spatially-invariant systems in Sections~\ref{sec:firstOrder}-\ref{sec:backstepping}. 
Extensions to spatially-varying systems are presented in Section~\ref{sec:spatiallyVarying}. Illustrative examples are provided in Section~\ref{sec:applications}: consensus of first-order subsystems and vehicular platoons. Section~\ref{sec:structured} analyzes the structure of the resulting controller implementation for the consensus problem.

A preliminary version of some of these results is reported in~\cite{jensen2020Backstepping}. The preliminary version does not include proofs, the analysis of the structure of controller implementations, results applying to the spatially-varying setting, as well as analytical results for the consensus application included in the present manuscript.

 \section{Notation \& Preliminaries} \label{sec:notation}

 We let $\mathcal{R}$ denote the set of proper and rational (possibly matrix-valued) transfer functions and let $\overline{\mathcal{R}}\subset \mathcal{R}$ denote the subspace of strictly proper transfer functions. We define a transfer function to be stable if it has no poles in the closed right half plane $\{ s: {\rm Re}(s) \ge 0\}$ and let $\mathcal{R}_s$ and $\overline{\mathcal{R}}_s$ denote the subsets of stable elements of $\mathcal{R}$ and $\overline{\mathcal{R}}$ respectively.
 The $\h_2$ norm provides one measure of a system $G \in\overline{\mathcal{R}}_s$:
 	$$
		 \| G\|_{\mathcal{H}_2}^2 : = {\rm trace}\left( \int_{- \infty}^{\infty} G^*(j\omega) G(j \omega) d\omega \right).
	$$

We consider spatially distributed dynamical (i.e. spatio-temporal)  systems where the state and all external signals are functions of a spatial variable $n$ as well as time $t$. We denote such spatio-temporal signals using lower case letters and denote their (temporal) Laplace transform using uppercase letters, e.g. the (possibly vector-valued) state at location $n$ and time $t$ is denoted by
\[
	x(n,t)  ~~\mbox{or}~~  x_n(t) , ~~~~n\in\cG, ~t\in\R^+ \!:=[0,\infty), 
\]
where the spatial index $n$  takes values in the finite set $\cG = \Z_N$ or the countably infinite set $\cG = \Z$, and its (temporal) Laplace transform is denoted by
	$$
		X_n(s)  ~~\mbox{or}~~  X(n,s) ~:=~ \int_0^\infty x_n(t) ~e^{-st} dt.
	$$
We also represent signals as finite or infinite vectors 
\begin{align*}
	x(t) &= \left[ \begin{array}{ccc} x_0(t) & \cdots & x_{N\sm1}(t) \end{array} \right]^T , 				\\
	x(t) &= \bbm \cdots  &  x_{\sm1}^T(t) &  x_0^T(t) &  x_1^T(t) &  \cdots		\ebm^T ,
\end{align*}
depending on whether $\cG$ is finite or countably infinite and similarly represent Laplace transforms of such signals $X(s)$.

We use the  $L^2(\G \times   \R^+)$  (denoted  simply as $L^2$) norm on (vector-valued)
 spatio-temporal signals, defined by
 	\begin{subequations} \begin{align}
		\|x\|^2_2&:=  \sum_{n \in \cG} \int_{0}^{\infty} x_n^*(t) x_n(t)~ dt \label{eq:L2time} \\
			&= \sum_{n \in \cG} \int_{-\infty}^{\infty} X_n^*(j\omega) X_n(j\omega)~ d\omega,
	\end{align} \end{subequations}
where $(^*)$ denotes the complex conjugate transpose, and the equality follows from the Plancherel Theorem.

\subsection{Spatially-Invariant Systems}
Transfer matrices are used to  describe LTI systems in the  finite space setting $\cG = \Z_N$, i.e. 
	\begin{equation*} 
		X(s) = H(s) ~U(s),  
	\end{equation*}
where $X(s) = \left[ \begin{array}{ccc} X^T_0(s) & \cdots &X^T_{N-1}(s) \end{array} \right]^T$ and $U(s) = \left[ \begin{array}{ccc} U^T_0(s) & \cdots &U^T_{N-1}(s) \end{array} \right]^T$ are the block partitioned input and output of the system $H$. In the special case that the transfer matrix $H$ is circulant, i.e.
	\begin{equation} \label{eq:Kcirculant}X(s)= \left[ \begin{array}{cccc} H_0(s) & H_{N-1}(s) & \cdots & H_1(s) \\ H_1(s) & H_0(s) & \cdots & H_2(s) \\ \vdots & & \ddots & \\ H_{N-1}(s) & H_{N-2}(s) & \cdots & H_0(s) \end{array} \right] U(s),
	\end{equation}
we refer to the system as \emph{spatially-invariant}. Relation \eqref{eq:Kcirculant} is written equivalently as a spatial convolution in the transfer function domain:
		\begin{equation} \begin{aligned} \label{eq:spatiallyInvSysRepresentation}
		X_n(s) = ({H} U)_n(s) &:=\big(H\left(s\right) * U(s)\big)_n\\
		&:= \sum_{m \in \G} H_m(s) U_{n -m}(s),
		\end{aligned} \end{equation}
where $(*)$ denotes circular convolution over $\cG = \Z_N$. With some abuse of notation, we also use $(*)$ to denote standard discrete convolution over $\cG = \Z$ so that representation \eqref{eq:spatiallyInvSysRepresentation} applies to the infinite spatial domain setting as well. In this infinite extent case ($\cG = \Z$), $H$ can be represented as an infinite extent Toeplitz transfer matrix, i.e.
	\begin{equation} \label{eq:KToeplitz}
	X(s) =  \left[ \begin{array}{cccccc} \ddots & \ddots & \ddots  \\
	 & H_1(s) & H_0(s) & H_{-1}(s) &  \\
	& & H_1(s) & H_0(s) & H_{-1} (s)\\
	 & & \ddots & \ddots & \ddots  
	 \end{array} \right] U(s),
	 \end{equation}
with each $H_m(s)$ a finite-dimensional transfer matrix. Note that for both $\cG = \Z_N$ and $\cG = \Z$, a spatially-invariant system $H$ is completely specified by the (possibly infinite) sequence of transfer functions $\{H_m(s)\}_{m \in \cG}$, which we refer to as the \emph{convolution kernel} of $H$. 
We say $H= \{ H_m(s)\}_{m \in \cG} \in \cR$ (resp. $\cR_s,~\oR,~\oRs$) if each element of the convolution kernel $H_m \in \cR$ (resp. $\cR_s,~\oR,~\oRs$). Spatial invariance is needed to formalize our infinite-dimensional results, which are studied to provide insight to the finite space setting as the number of subsystems $N \rightarrow \infty$. 
	Throughout the remainder of the paper we write \emph{spatially-invariant system} to refer to both finite and infinite extent settings unless otherwise stated.
Special classes of spatially-invariant systems include
	\begin{itemize}
	\item 
		$B = \{ B_n\}_{n \in \G}$ is a \emph{pointwise multiplication operator} if $B_n= 0$ for all $n \ne 0$. With some abuse of notation we often denote $B_0 = B$. 	
		\item 
	$C$ is a \emph{spatial convolution operator} if it is of the form 
		\begin{equation*} 
		({C} x)_n(t) :=(C * x)_n(t) = \sum_{m \in \G}  C_m x_{n-m}(t),  
		\end{equation*}
	with $\{C_m\}_{m \in \cG}$ a sequence of {scalar-valued} matrices. 
	\end{itemize}
Pointwise multiplication operators are represented by block diagonal (possibly infinite extent) matrices and spatial convolution operators are represented by {static} circulant (or Toeplitz) matrices.

\begin{exmp}
	The \emph{temporal differentiation operator}, 
		$$ \dot x_n(t) := \frac{d}{d t} x_n(t),$$
is a pointwise multiplication operator. It can be represented in the transfer function domain as multiplication by the diagonal (potentially infinite-dimensional) matrix $sI$:
		$$\dot{x}_n(t) ~ \leftrightarrow sI \cdot X_n(s).$$
\end{exmp}

\subsection{Locality Constraints}
We are interested in the design of \emph{localized} controllers. In other words, the control action at each spatial site is computed using only information from \emph{nearby} spatial locations where `nearby' is specified by a known underlying communication graph. We formalize this notion as follows. 

	\begin{defn} \label{def:structure_moregeneral}
	Let $H$ be a transfer matrix (or real-valued matrix), block partitioned as
	$$H = \left[ \begin{array}{cccc} H_{0,0} & H_{0,1}& \cdots & H_{0,N-1} \\ H_{1,0} & H_{1,1} & \cdots & H_{1,N-1} \\ \vdots & \\ H_{N-1,0} & H_{N-1,1} & \cdots & H_{N-1,N-1} \end{array} \right],$$
	and let $\mathcal{A} \in \R^{N \times N}$ denote an adjacency matrix of a communication graph with $N$ nodes. We say that $H$ is \emph{structured} with respect to $\mathcal{A}$ if $H_{ij} = 0$ whenever $\mathcal{A}_{ij} = 0$. 
	\end{defn} 
	
	To specialize this notion to the spatially-invariant setting, we consider a circle graph $(\cG = \Z_N)$ or an infinite chain graph $(\cG = \Z)$ with edges between nodes of distance $\le M$ away\footnote{in the case of the circle this distance is computed modulo the number of nodes} and denote the corresponding adjacency matrix by $\mathcal{A}^M$.
	
	\begin{defn} \label{def:bandsize}
	The spatially-invariant system $H$ is \emph{structured} w.r.t. $ \mathcal{A}^M$ if its representation \eqref{eq:spatiallyInvSysRepresentation} can be written as
	$$
		X_n(s)=\sum_{|m| \le M} H_m(s) U_{n-m}(s),
	$$
	i.e. the convolution kernel defining $H$ has entries $H_n(s) \equiv0$ for $|n| > M$. We refer to $M$ as the \emph{band size} of $H$.
	\end{defn}

 Note that the definitions of structure and band size of a system depend only on its {transfer function} representation, and are independent of the chosen {state-space realization.}

We summarize some useful properties of spatially-invariant systems in the following proposition; the proof follows immediately from the representation \eqref{eq:spatiallyInvSysRepresentation} and is omitted. We refer the reader to \cite{bamieh2002distributed, curtain2009system} for a review of this class of systems.
	\begin{prop} \label{prop:spatiallyInvProperties}
	Let $K$ and $H$ be spatially-invariant systems. Assume $K, H$ are defined on signal spaces of appropriate dimensions so that the composition operator
	$(KH)(x) = K(H(x))$
	is well-defined. Then the following hold:
	\begin{enumerate}
	\item $KH$ is a spatially-invariant system. In particular, for each positive integer $n$, $K^n$ denotes the composition of $K$ $n$ times and is a spatially-invariant system. 
	\item If $K$ and $H$ have finite band sizes $M$ and $L$, respectively, then $KH$ has band size $L+M$. Thus the composition of two spatially-invariant systems, each with finite band size, will also have finite band size.
	\item The inverse operator $K^{-1}$, when it exists, is also a spatially-invariant system. 
	\end{enumerate}
	\end{prop}

\section{Problem Formulation} \label{sec:setup}
We consider the design of a (dynamic or static) 
state-feedback controller
	 \begin{equation} \label{eq:stateFB}
	 	u(t)  = (Kx)(t),
	\end{equation}
for a spatially-distributed plant $P$ with dynamics
	\begin{subequations} \label{eq:spatiallyInvGeneral} \begin{align} \label{eq:spatiallyInv_state}
	\dot{x}(t) &= (Ax)(t) + (B_1 w)(t) + (B_2 u)(t) \\
	z(t) &= (C_1 x)(t) + (D_{12} u)(t), \label{eq:spatiallyInv_perf}
	\end{align} \end{subequations}
where the spatio-temporal signals $x, u,$ $w$, and $z$ represent the spatially distributed state, control action and exogenous disturbance respectively. Recall that e.g. $x_n(t)$ represents the local state at spatial index $n \in \G$ at time $t \in \R^+$.  
We consider the following two settings:
	\begin{enumerate} [label=\roman*)]
		\item $A, B_1, B_2, C_1, D_{12}$ are finite-dimensional real-valued matrices and $K$ can be represented by a  transfer function with finitely many inputs/outputs\footnote{With some abuse of notation we denote by $K$ the mapping \eqref{eq:stateFB} or the corresponding transfer function}, or 
		\item $A, B_1, B_2, C_1, D_{12}$ are spatial convolution operators and $K$ is a spatially-invariant system. The spatial domain in this case may be finite ($\cG = \Z_N$) or infinite $(\cG= \Z)$.
	\end{enumerate}

 \subsection{Closed-Loop Mappings}
  Consider the system \eqref{eq:spatiallyInvGeneral} in feedback with the (static or dynamic) state-feedback controller \eqref{eq:stateFB}.
We let $\Phi^x$ and $\Phi^u$ denote the resulting \emph{closed-loop mappings} from disturbance $B_1w$ to state $x$ and control action $u$, respectively. In the transfer function domain \cite{wang2019system, jensen2018optimal}
	\begin{equation} \begin{aligned} \label{eq:phi_inf}
	\left[ \begin{array}{c}X \\ U \end{array} \right] &= \left[ \begin{array}{c}\Phi^x \\ \Phi^u \end{array} \right] B_1 W \\
	&:= \left[ \begin{array}{c}(sI - A - B_2 K)^{-1} \\ K(sI - A - B_2 K)^{-1}   \end{array} \right] B_1 W.
	\end{aligned} \end{equation} 
In either case (i) or case (ii), internal stability of the feedback interconnection of plant $P$ with controller $K$ can be derived from the stability of these closed-loop mappings. Formally, 
$K$ is \emph{(internally) stabilizing} for \eqref{eq:spatiallyInvGeneral} if the resulting closed-loop mappings $\Phi^x, ~\Phi^u$ are elements of $\oR_s$. Moreover, $K$ can be recovered from these closed-loops as 
	$$ u = K x := \Phi^u \left( \Phi^x \right)^{-1} x.$$
The closed-loop map from disturbance $w$ to performance output $z$ can be written in terms of these closed-loops as 
	\begin{equation} \begin{aligned}
		z & = \F(P;K) w := \left[ \begin{array}{cc} C_1 & D_{12} \end{array} \right] \left[ \begin{array}{c}\Phi^x \\ \Phi^u \end{array} \right] B_1 w
	\end{aligned} \end{equation}

In case (i), $\Phi^x(s)$ and $ \Phi^u(s)$ are transfer matrices with finitely many inputs and outputs. When $P$ is spatially-invariant  (case (ii)), we impose the additional constraint that 
		$K$ is spatially-invariant. In this case, the following proposition demonstrates that $\Phi^x$ and $\Phi^u$ are spatially-invariant systems; the proof of this result follows directly from Proposition \ref{prop:spatiallyInvProperties}.
	\begin{prop} \label{prop:KandPhi}
	Let $u(t)=(Kx)(t)$ be a controller for \eqref{eq:spatiallyInvGeneral} with $A, B_1,$ and $B_2$ spatial convolution operators. Then $K$ is a spatially-invariant system if and only if $\Phi^x, \Phi^u$ \eqref{eq:phi_inf} are spatially-invariant. In this case, the closed-loop map $\mathcal{F}(P;K)$ is spatially-invariant as well.
	\end{prop}

\subsection{Optimal Controller Design}

We design the controller \eqref{eq:stateFB} to optimize some closed-loop performance measure and, following the System Level Approach \cite{anderson2019system}, impose additional locality constraints on the {closed-loop mappings} $\Phi^x$ and $\Phi^u$. We focus on the setting of spatially-invariant systems; in this case the controller design problem of interest is as follows.\footnote{
When $P$ is not spatially-invariant, the spatial invariance constraint on $K$ is removed from the optimal controller design problem.
}
\\
 \textbf{Optimal spatially-invariant controller design with closed-loop spatial sparsity constraints:}
	\begin{equation} \label{eq:optAnyNorm}
	\begin{aligned} 
	&\inf_{K \text{stabilizing}} ~ \left\| \left[ \begin{array}{cc} C_1 & D_{12} \end{array} \right] \left[ \begin{array}{c}\Phi^x \\ \Phi^u \end{array} \right] B_1 \right\|\\
	&~~~\text{ s.t. }~~~~ \Phi^u, \Phi^x \text{ have band size }M\\
	& ~~~~~~~~~~~~K \text{ spatially-invariant} 
	\end{aligned}
	\end{equation}
where $\| \cdot \|$ denotes any well-defined system norm; common choices (in the finite-dimensional setting) include the $\h_2, \h_{\infty},$ or $L_1$ norm.

We note that the closed-loop structured controller design problem \eqref{eq:optAnyNorm} is \emph{not} the same as the design problem subject to constraints on the controller itself. However imposing this closed-loop band size constraint
\footnote
		{
			For simplicity we restrict to an odd number of subsystems $N$ in the finite space setting so that these constraints will be symmetric.
		}
		 preserves convexity and has the following consequences: 
\begin{enumerate}
\item If $\Phi^u$ and $\Phi^x$ have band size $M$, then the corresponding controller has an implementation for which subcontroller communication is limited to a neighborhood of size $M$ (Sec.~\ref{sec:structured}),
\item When $\| \cdot\|$ denotes the $\h_2 $ norm, \eqref{eq:optAnyNorm} can be converted to a standard \emph{finite-dimensional} model-matching problem with $2M+1$ transfer function parameters where $M$ is the constrained closed-loop band size. This holds even in the infinite spatial extent spatially-invariant setting, provided $C_1$ and $D_{12}$ have finite band size (Sec.~\ref{subsec:modelmatch}).
\end{enumerate}

\subsection{Decoupled Dynamics}

We focus on spatially distributed plants that satisfy the assumption
\begin{enumerate}
\item \textbf{(Decoupled subsystem dynamics)} The state equation \eqref{eq:spatiallyInv_state} for $P$ can be written as:
	\begin{equation}  \label{eq:decoupledDynamics}
	\dot{x}_n(t) = A^{(n)} x_n(t) + B_{1}^{(n)} w_n(t) + B_{2}^{(n)}u_n(t),~ n \in \G,
	\end{equation} 
	i.e. the open-loop dynamics of subsystem $n$ are independent of all other subsystems $m \ne n$. 
	(Note that coupling may be introduced in the performance output by the operators $C_1$ and $D_{12}$.)
\end{enumerate}

In Appendix~\ref{app:coupled} we provide analogous results for systems with \emph{coupled} subsystem dynamics, allowing for $A$ and $B_2$ to be spatial convolution operators (with finite band size) which additionally satisfy either
	\begin{enumerate}[label=\alph*)]
		\item The open-loop dynamics are {stable}, or
		\item The operator from control to state $(B_2)$ is invertible. 
	\end{enumerate}
Results presented in Appendix~\ref{app:coupled} follow from ideas of \cite{jensen2018optimal}. Extensions to general distributed systems with coupled subsystem dynamics (i.e. removing assumptions (a) or (b)) remains an open problem and is the subject of future work.

	Distributed systems composed of subsystems with decoupled open-loop dynamics, i.e. systems satisfying assumption (1), include the following examples which will be analyzed in more detail in Section~\ref{sec:applications}.
	\begin{enumerate}
\item \textbf{Consensus of first-order subsystems:}
	The dynamics are given by
	\begin{equation} \label{eq:1storder}
	\dot{x}_n = u_n + w_n,~~ n \in \G,
	\end{equation}
	with each $x_n(t), u_n(t), w_n(t)$ scalar-valued. Applications for the first-order consensus problem include load balancing over a
distributed file system.
\item \textbf{Vehicular platoons (consensus of second order subsystems):}
Following \cite{jovanovic2005ill}, we let $\xi_n$ represent the absolute deviation of vehicle $n$ from a desired trajectory $\overline{\xi}_n$
	$$\overline{\xi}_n := \overline{v}t + n \delta,$$
with $\overline{v}$ the specified cruising velocity.  
Defining $v_n: = \dot{\xi}_n$, the dynamics are given by
	\begin{equation} \begin{aligned} \label{eq:2ndorder_standard}
	\left[ \begin{array}{c} \dot{\xi}_n \\ \dot{v}_n \end{array} \right] = \left[ \begin{array}{cc} 0 & 1 \\ 0 & 0 \end{array} \right]  \left[ \begin{array}{c} {\xi}_n \\ {v}_n \end{array} \right]  + \left[ \begin{array}{c} 0 \\ 1 \end{array} \right] \left(u_n + w_n \right),n \in \G.
	\end{aligned} \end{equation}
\end{enumerate}

The setting of spatially-invariant systems (case (ii)) with decoupled subsystem dynamics is analyzed in detail in Sections~\ref{sec:firstOrder} -\ref{sec:backstepping}. 
In Section \ref{sec:spatiallyVarying}, we will analyze decoupled subsystem dynamics that are \emph{spatially varying} (case (i)).

\section{Closed-Loop Parametrization: Locally 1\textsuperscript{st} Order Systems} \label{sec:firstOrder}

In this section, we study systems with dynamics of the form:
	\begin{equation}\label{eq:1stSpatiallyInv}
	 \dot{x}_n = a x_n + w_n + u_n, ~ n \in \G.
	 \end{equation}
In vector form \eqref{eq:1stSpatiallyInv} is written as
\begin{equation} \label{eq:vectorForm}
	\dot{x}  ~=~ A x ~+~ w ~+~ u , 
\end{equation}
the ``A-matrix'', $A := aI$, is a multiple of the (possibly infinite extent) identity matrix. The dimension of the overall state vector $x(t)$ is equal to the cardinality of $\cG$, and can be either finite or infinite. We call this class of systems \emph{spatially-invariant locally $\first$ order.} The first-order consensus problem \eqref{eq:1storder} provides one application of this system class; this example will be further analyzed in Section~\ref{sec:1stOrder}.

The following theorem provides an explicit parameterization of all achievable stabilized closed-loop mappings for spatially-invariant locally $\first$ order systems in the case that ${\rm Re}(a) \ge 0$ (unstable plant dynamics)\footnote
	{
		The parameterizations presented in Appendix~\ref{app:coupled} apply to the stable open-loop setting $(a < 0)$.
	}.
	\begin{thm} \label{thm:infFirstOrder}
	 Let ${\rm Re}(p) >0$. 
	A spatially-invariant state-feedback controller $K$ is stabilizing for the spatially-invariant locally $\first$ order system \eqref{eq:1stSpatiallyInv} with ${\rm Re}(a) \ge 0$ if and only if the resulting closed-loop mappings are of the form 
	\begin{subequations} \label{eq:infFirstOrder_XU} \begin{align}
	 \label{eq:infFirstOrderX}
	\Phi^x &= (sI + pI )^{-1} \theta + (sI + pI )^{-1} \\
	\Phi^u&= (s-a)(sI+pI)^{-1} \theta - (a+p)(sI+pI)^{-1}\label{eq:infFirstOrderU}
	\end{align} \end{subequations}
	for some stable, spatially-invariant $\theta= \{ \theta_n(s)\}_{n \in \cG} \in \oRs$. 
	\end{thm}
	
\begin{rem}
	An alternate parameterization for the class of locally first-order systems is provided in Appendix~\ref{app:coupled}.
The usefulness of parameterization \eqref{eq:infFirstOrder_XU} over this alternate is that it may be modified to apply to systems with operator $B_2$ not invertible. This modification is provided in Section~\ref{sec:backstepping}.
\end{rem}

\begin{IEEEproof}
We consider here the finite space setting $\cG = \Z_N$;  details for the infinite space setting are presented in Appendix~\ref{app:proofs}.   By Proposition~\ref{prop:spatiallyInvProperties}, $K$ is spatially-invariant if and only if $\Phi^x$ and $\Phi^u$ are.  Then following the results of \cite{wang2019system}, the set of all closed-loop mappings resulting from the feedback interconnection of \eqref{eq:1stSpatiallyInv} with an internally stabilizing spatially-invariant controller $u = Kx$ is implicitly parameterized by the affine subspace: 
	\be \label{eq:proof41}
		\begin{cases} 
		(sI-aI) \Phi^x(s) - \Phi^u(s) = I\\
		\Phi^x, \Phi^u \in \oRs, ~ \text{spatially-invariant}
		\end{cases}
	\ee
Thus, it is sufficient to show that the resulting closed-loop mappings $\Phi^x, \Phi^u$ for \eqref{eq:1stSpatiallyInv} are in the affine subspace \eqref{eq:proof41} if and only if they are of the form \eqref{eq:infFirstOrder_XU} for a spatially-invariant $\theta \in \oRs$. 

First assume $\Phi^x, \Phi^u$ are of the form \eqref{eq:infFirstOrder_XU} for a spatially-invariant $\theta \in \oRs$. Then by Proposition~\ref{prop:spatiallyInvProperties}, $\Phi^x$ and $\Phi^u$ are spatially-invariant. Direct computations show that $(sI-aI) \Phi^x(s) - \Phi^u(s) = I$ and that $\Phi^x, \Phi^u$ are strictly proper. Stability of $\theta$ and $(sI - pI)^{-1}$ imply stability of $\Phi^x$ and $\Phi^u$. 

Next assume $\Phi^x, \Phi^u$ are in the subspace defined by \eqref{eq:proof41}. Then 
	\be
		I + \Phi^u(s) = (s-a) \Phi^x(s), ~~\Phi^x \in \oRs
	\ee
implies the interpolation constraint
	\be
		\Phi^u(a) = -I
	\ee
must hold. Equivalently, $\Phi^u$ can be decomposed into the sum of two terms as 
	\be \label{eq:proof41_u}
		\Phi^u(s) = \frac{s-a}{s+p} \theta(s)- \frac{a+p}{s+p}I, 
	\ee
Substituting this into the relation $\Phi^x(s) = \frac{1}{s-a} \left( I + \Phi^u(s) \right)$, we see that 
	\be \label{eq:proof41_x}
		\Phi^x(s) = \frac{1}{s+p} \theta(s) + \frac{1}{s+p}I.
	\ee
It remains to show that $\theta\in \oRs$. Rearranging \eqref{eq:proof41_x}, we see that stability of $\Phi^x$ implies 
	$$\theta(s) = (s+p) \left( \Phi^x(s) - I \right)$$ is stable. Rearranging \eqref{eq:proof41_u}, strict properness of $\Phi^u$ implies
	$$\theta(s) = \frac{s+p}{s-a} \Phi^u(s) + \frac{a+p}{s-a}I$$
is the sum of two strictly proper transfer matrices, and is therefore itself strictly proper. 
\end{IEEEproof}

Theorem~\ref{thm:infFirstOrder} shows that each stable and strictly proper $\theta$ leads to a stabilizing controller $K$. Thus, the optimal controller design problem \eqref{eq:optAnyNorm} may be written with this parameter $\theta \in \oRs$ as a decision variable; this will be formalized in Equation~\eqref{eq:almostModelMatching}. Theorem~\ref{thm:infFirstOrder} applies to a quite restrictive setting but will be generalized to spatially-invariant systems with higher order subsystem dynamics in Section~\ref{sec:backstepping} and to the spatially-varying setting in Section~\ref{sec:spatiallyVarying}.

Note that like the standard Youla parameterization, \eqref{eq:infFirstOrder_XU} provides an explicit parameterization of all stabilizing controllers. Thus, we claim that the co-prime factorization of the plant that must be explicitly computed for the standard Youla parameterization appears to be built in to our parameterization procedure. This is illustrated (for the setting of identical unstable first-order subsystems) in the following section, which demonstrates the relation of our parameter $\theta$ to the standard Youla parameter $Q$. We remark that \cite{zheng2020equivalence} has also made comparisons between Youla and alternate controller parameterizations.

\subsection{Relation to Youla Parameterization} \label{sec:youla_connection}

First note that a general open-loop mapping from control $u$ to state $x$ of the form $(sI-A)^{-1}B$ can be factored as:
	\be
		(sI-A)^{-1}B = \left( I - \frac{1}{s+p} (A + pI) \right)^{-1} \frac{1}{s+p} B 
	\ee
For any ${\rm Re}(p)>0$ this is in fact a left coprime factorization if and only if the pair $(A,B)$ is stabilizable; to see this note that 
	\be
		{\rm rank} \left[\begin{array}{cc}I - \frac{1}{s+p} (A + pI) & \frac{1}{s+p} B   \end{array} \right] = {\rm rank}  \left[ \begin{array}{cc} sI - A & B \end{array} \right].
	\ee

For system \eqref{eq:1stSpatiallyInv} we denote
the open-loop mappings from control $u$ to measurement $x$ by $P_{22}$; by stabilizability of $(A, B) = (aI, I)$, a left co-prime factorization is given by 
	\be \begin{aligned}
		P_{22} = (sI - aI)^{-1} &= \left( I - \frac{a+p}{s+p}\right)^{-1} \frac{1}{s+p}I\\
		& = \left(\frac{s-a}{s+p}I \right)^{-1} \frac{1}{s+p}I =: M_{\ell}^{-1} N_{\ell},
	\end{aligned} \ee
for ${\rm Re}(p) >0$.  In this case, the factors commute and we immediately obtain a right co-prime factorization as well: 
	\be \begin{aligned}
		P_{22} = (sI - aI)^{-1} &= \frac{1}{s+p} I \left(\frac{s-a}{s+p} I\right)^{-1} =: N_r M_r^{-1}.
	\end{aligned} \ee
The set of all stabilizing controllers $K$ is then parameterized using the 	
corresponding Bezout identity as follows: $K$ is a stabilizing controller for \eqref{eq:1stSpatiallyInv} if and only if 
	\be \label{eq:K_coprime}
		K = (V_r - M_r Q)(U_r - N_r Q)^{-1}
	\ee
for some stable  $Q$ \cite{rotkowitz2005characterization}, where $U_r, V_r, U_{\ell}, V_{\ell}$ are given by 
	\begin{equation*} \begin{aligned}
		I ~&~ =
	\left[ \begin{array}{cc} U_{\ell} & - V_{\ell} \\ - N_{\ell} & M_{\ell} \end{array} \right] \left[ \begin{array}{cc} M_r & V_r \\ N_r & U_r \end{array} \right]\\
		&~ =\left[ \begin{array}{cc}  \frac{s + 2p + a}{s+p}I &  \frac{(p+a)^2}{s+p}I \\ \frac{-1}{s+p}I & \frac{s-a}{s+p}I \end{array} \right]
		\left[ \begin{array}{cc}  \frac{s-a}{s+p}I  &  \frac{-(p+a)^2}{s+p} \\ \frac{1}{s+p}I &  \frac{s + 2p + a}{s+p}I \end{array} \right].
	\end{aligned} \end{equation*}
Moreover, $K$ will be spatially-invariant if and only if $Q$ is. The closed-loop mappings for \eqref{eq:1stSpatiallyInv} corresponding to the controller \eqref{eq:K_coprime} are given by 
	\begin{equation} \begin{aligned}	\label{eq:Phix_Q}
		\Phi^x & = (sI - aI - K)^{-1} \\
		&= (sI - aI -  (V_r - M_r Q)(U_r - N_r Q)^{-1})^{-1}\\
		& = \frac{s+2p+a} {(s+p)^2} I - \frac{1}{(s+p)^2} Q, 
	\end{aligned} \end{equation}
	\begin{equation} \begin{aligned} 	\label{eq:Phiu_Q}
		\Phi^u & = K \Phi^x
		 = (V_r - M_r Q)(U_r - N_r Q)^{-1} \Phi^x\\
		 &=\frac{-(a+p)^2}{(s+p)^2}I - \frac{s-a}{(s+p)^2 }Q
	\end{aligned} \end{equation}
Equating parameterizations \eqref{eq:Phix_Q}-\eqref{eq:Phiu_Q} to the parameterizations \eqref{eq:infFirstOrderX} of Theorem~\ref{thm:infFirstOrder}, the parameter $\theta$ of \eqref{eq:infFirstOrderX} can be written in terms of the standard Youla parameter $Q$ as
	\be 
		\theta = \frac{a+p}{s+p} I - \frac{1}{s+p} Q.
	\ee

\section{Locally $n$\textsuperscript{th} Order Systems: A Back-Stepping Approach} \label{sec:backstepping}
To generalize the results of Section \ref{sec:firstOrder} to the case of subsystems with higher order dynamics, 
we begin by considering a single finite-dimensional system with dynamics of the form 
	\begin{equation} \label{eq:nthOrder}
	\dot{x} ~=~ A x ~+~ B_1 w ~+~ B_2 u,
	\end{equation}
with $(A, B_2)$ controllable. We assume that $(A+I, B_2)$ is in \emph{controllable-canonical form} \cite{hespanha2018linear}, i.e.
	\begin{equation} \begin{aligned} \label{eq:ControllableCanonical}
	({A}+I)&= \left[ \begin{array}{ccccc} -a_1 I_m & -a_2 I_m & -a_3I_m & \cdots & -a_{r} I_m\\
						      I_m & 0 & 0 & \cdots & 0 \\ 
						      0 & I_m & 0 & \cdots & 0 \\
						      \vdots & & \ddots & \ddots  \\
						      0 & & & I_m & 0 
						       \end{array} \right],\\
	{B}_2&= \left[ \begin{array}{ccccc} I_m & 0 & 0 & \cdots & 0\end{array} \right]^T,
	\end{aligned} \end{equation}
where $I_m$ denotes the $m \times m$ identity matrix. $r$ is defined to be the order of the system. We note that the assumed form \eqref{eq:ControllableCanonical} is nonstandard and is chosen to simplify the use of Lemma \ref{lem:newAffineConstraint}. Moreover, this form may be assumed without loss of generality, as demonstrated by the following proposition. 
 	\begin{prop} \label{prop:controllableCanonical}
	If the finite-dimensional system $(A, B_2)$ is controllable, then there exists an invertible transformation matrix $T$ such that $\hat{A}:= T A T^{-1}, \hat{B}_2 := TB_2$ and $( \hat{A} + I, \hat{B}_2 )$ is in {controllable-canonical form}, i.e. of the form  \eqref{eq:ControllableCanonical}.
	\end{prop}
	\begin{IEEEproof} $A$ and $(A+I)$ have the same set of eigenvectors, so that $(A,B_2)$ is controllable if and only if $(A+I,B_2)$ is. Since $(A+I,B_2)$ is controllable, there exists a similarity transformation which converts the system to controllable canonical form \cite{hespanha2018linear}. 
	\end{IEEEproof}

We remark that this state transformation may be computationally expensive for very high order systems. However, many relevant applications for this procedure are distributed systems where the order of each individual subsystem is small; the state transformation can be computed for each (low order) subsystem and is therefore relatively inexpensive to compute. 
	\begin{exmp} \label{exmp:vehicleDynamics}
		Each subsystem of the vehicular platoons problem is of order $r = 2$. Defining a new state, $x_n:= \left[ \begin{array}{cc} ({v}_n+
	 {\xi}_n) & {\xi}_n \end{array} \right]^T$, the dynamics \eqref{eq:2ndorder_standard} for each order 2 subsystem may be written in form \eqref{eq:ControllableCanonical} as
	\begin{equation} \begin{aligned} \label{eq:2ndorder}
	\dot{x}_n= \left[ \begin{array}{cc} 1 & -1 \\ 1 & -1 \end{array} \right]  x_n + \left[ \begin{array}{c} 1 \\ 0 \end{array} \right] w_n + \left[ \begin{array}{c} 1 \\ 0 \end{array} \right] u_n,~ n \in \G.
	\end{aligned} \end{equation}
	The dynamics of the overall system can be written as 
		\begin{equation}
			\dot{x} = A x + B_1 w + B_2 u, 
		\end{equation}
	where $A, B_1, $ and $B_2$ are the operators of pointwise multiplication by the matrices 
		\begin{equation} 
			A:=  \left[ \begin{array}{cc} 1 & -1 \\ 1 & -1 \end{array} \right] , ~~ B_1 = B_2 = \left[ \begin{array}{c} 1 \\ 0 \end{array} \right].
		\end{equation}
		This application will be analyzed in further detail in Section~\ref{sec:2ndOrder}. 
	\end{exmp} 
	
\begin{rem}
	When the original system is not in the form \eqref{eq:nthOrder} with $(A+I, B_2)$ controllable canonical, the controller is designed in terms of the transformed state $z:= Tx$ as $u = \hat{K} z$ and then transformed back to the original coordinates as $u = (\hat{K} T )x =: K x$. The state transformation for open-loop decoupled dynamics may be done site by site, so that these transformations preserve any locality structure. 
\end{rem}

To extend our parameterizations to the case of higher order subsystems we employ the following lemma, which follows from a straightforward modification of \cite[Thm. 1]{wang2019system}. The reasoning behind this modification is that for any ${\rm Re}(p)>0$ operation of multiplication by $(sI+pI)$ has an inverse that preserves stability, i.e. $(sI + pI)^{-1} \theta \in \oRs$ whenever $\theta \in \oRs$. 

\begin{lem} \label{lem:newAffineConstraint}
	There exists a state feedback controller for \eqref{eq:nthOrder} that results in the closed-loop mappings $\Phi^x, \Phi^u$ if and only if $\Phi^x, \Phi^u$ are strictly proper and satisfy the affine constraint 
	\begin{equation} \label{eq:newAffineConstraint}
		\left[ \begin{array}{cc} (sI+pI)-(A+pI) & -B_2 \end{array} \right] \left[ \begin{array}{c} \Phi^x(s) \\ \Phi^u(s)  \end{array} \right] = I,
	\end{equation}
where $p$ is any scalar.
	If $\Phi^x$ and $\Phi^u$ are stable, i.e. $\Phi^x, \Phi^u \in \oRs$, then the corresponding controller is stabilizing.
\end{lem}

Analogous to \cite[Lem. 1]{wang2019system}, existence of a solution $ \left[ \begin{array}{c} \Phi^x \\ \Phi^u  \end{array} \right] \in \oRs$ to \eqref{eq:newAffineConstraint} is equivalent to stabilizability of $(A,B_2)$. 
We employ Lemma \ref{lem:newAffineConstraint} with parameter $p=1$ to prove the following theorem\footnote{
The choice of pole at $-p = -1$ is arbitrary, and analogous formulations would hold for alternate choice of stable pole $-p$ with $(A+pI, B_2)$ in controllable-canonical form
}.

	\begin{thm} \label{thm:nthOrder} 
	The controller \eqref{eq:stateFB} is stabilizing for the system \eqref{eq:nthOrder} of order $r$, with $(A+I, B_2)$ in controllable-canonical form, if and only if the resulting closed-loop mappings are of the form
		\begin{equation} \begin{aligned}  \label{eq:Phi_nthOrder}
		\Phi^x(s)  &=  \left[ \begin{array}{c} \frac{1}{s+1}I_m \\ \frac{1}{(s+1)^2}I_m \\ \vdots \\ \frac{1}{(s+1)^r}I_m \end{array} \right]  \theta(s)~+\\
		&\small{ \left[ \begin{array}{ccccc} \frac{1}{s+1}I_m & 0 & 0& \cdots & 0\\
		\frac{1}{(s+1)^2}I_m & \frac{1}{s+1}I_m & 0&\cdots  & 0 \\
		\frac{1}{(s+1)^3}I_m & \frac{1}{(s+1)^2}I_m & \frac{1}{s+1}I_m &\cdots  &0 \\
		\vdots & & & \ddots \\
		\frac{1}{(s+1)^r}I_m & \frac{1}{(s+1)^{r-1}}I_m & \frac{1}{(s+1)^{r-2}}I_m &\cdots  & \frac{1}{s+1}I_m\end{array} \right]}\\
		~\\
		&=: F(s) ~ \theta(s) ~ +~ L(s), \\
		~\\
		\Phi^u(s) &= \chi(s)~ \theta(s)~ + ~\eta(s) 
		\end{aligned} \end{equation}
	for some $\theta \in \oRs$ of dimension $m \times mr$, where $$\chi(s):= 1 +\frac{a_1}{s+1} + \frac{a_2}{(s+1)^2} + \cdots + \frac{a_r}{(s+1)^{r}} $$
	with the $a_i$'s given by \eqref{eq:ControllableCanonical}
and $\eta(s)$ is a transfer function of dimension $m \times mr$ block partitioned as 
	\begin{equation*} \begin{aligned} 
		&\eta(s) =   \left[ \begin{array}{cccc} \eta^1(s) & \eta^2(s) & \cdots & \eta^r(s)\end{array} \right]\\
		&\eta^k(s):= \sum_{i=k}^{r} \frac{a_i}{(s+1)^{i + 1-k}}I.
	\end{aligned} \end{equation*}
	\end{thm}
\begin{IEEEproof} 
By Lemma \ref{lem:newAffineConstraint}, it is sufficient to show that $\Phi^x, \Phi^u \in \oRs$ satisfy the affine constraint \eqref{eq:newAffineConstraint} if and only if they are of the form \eqref{eq:Phi_nthOrder}. 
To show this we employ a backstepping-like procedure similar to the approach for strict feedback systems presented in \cite{krstic1995nonlinear}. Various works have employed similar techniques, e.g. \cite{jovanovic2005lyapunov, jovanovic2004architecture}. 
The full proof of Theorem \ref{thm:nthOrder} is presented in Appendix~\ref{app:proofs}. \end{IEEEproof}

\begin{rem}
Equations \eqref{eq:Phi_nthOrder} can be modified to provide a parameterization of all stabilized closed loops for {discrete-time} systems with dynamics of the form  
	$$
		x(t+1)~ = ~A x(t) + B_1 w(t) + B_2 u(t)
	$$
with $(A, B_2)$ in {standard} controllable-canonical form \cite{hespanha2018linear} by replacing all $(s+1)^{-1}$ terms with $z^{-1}$ and using the discrete-time definitions of $\Phi^x, \Phi^u$ presented in \cite{wang2019system}. \end{rem}

\subsection{Spatially-Invariant Locally $n^{\text{th}}$ Order Systems}
We next consider a spatially-invariant system composed of subsystems that each have dynamics of the form \eqref{eq:nthOrder}. Consider a system (of finite or infinite spatial extent)  with dynamics
	\begin{equation} \label{eq:infDimZ}
	\dot{x}_n ~=~ A x_n + B_1 w_n + B_2 u_n,~~ n \in \G,
	 \end{equation}
where $(A+I, B_2)$ is in controllable-canonical form. $r$ denotes the order of each subsystem so that $x_n = \left[ \begin{array}{cccc} x_{n1}^T & x_{n2}^T & \cdots & x_{nr}^T \end{array} \right]^T$ for each $n \in \cG$. We refer to systems of the form \eqref{eq:infDimZ} as \emph{spatially-invariant locally $r^{\text{th}}$ order}.

	\begin{thm} \label{thm:infDim} 
	A spatially-invariant state feedback controller $K$ is stabilizing for the spatially-invariant locally $r^{\text{th}}$ order plant \eqref{eq:infDimZ} if and only if the resulting closed-loops $\Phi^x= \{ \Phi^x_n\}_{n \in \G}$ and $\Phi^u=\{ \Phi^u_n \}_{n \in \G}$ are of the form:
	\begin{equation} \begin{aligned} \label{eq:nthOrderPhiComponents}
	\Phi^x_n(s) &= \begin{cases} F(s) \theta_n(s) + L(s),~ n =0\\ F(s) \theta_n(s),~ n \ne 0 \end{cases}\\
	\Phi^u_n(s) &= \begin{cases} \chi(s) \theta_n(s) + \eta(s),~n = 0\\\chi(s) \theta_n(s),~ n \ne 0, \end{cases}
	\end{aligned} \end{equation}
	for some spatially-invariant $\theta = \{ \theta_n\}_{n \in \G} \in \oRs$, where $F(s), L(s), \chi(s),$ and $\eta(s)$ are defined in Theorem \ref{thm:nthOrder}. Equivalently, 
	\begin{equation} \begin{aligned} \label{eq:nthOrderPhi}
	\Phi^x &= F \theta + L\\
	\Phi^u &= \chi \theta + \eta,
	\end{aligned} \end{equation}
	where $F, L, \chi, \eta$ are spatially-invariant systems defined by pointwise multiplication by $F(s), L(s), \chi(s), \eta(s)$. 
	\end{thm}	 
		 	\begin{IEEEproof} See Appendix~\ref{app:proofs}.	\end{IEEEproof}

These parameterizations are utilized to simplify the optimal controller design problem and easily allow for structural constraints on the closed loops.
From  \eqref{eq:nthOrderPhiComponents} we see that the closed-loop mappings $\Phi^x, \Phi^u$ will have the same band size as the parameter $\theta$. Thus the optimal spatially-invariant, closed-loop structured controller design problem for the class of systems considered in this section is given in terms of the parameter $\theta$ by
	\begin{equation} \begin{aligned} \label{eq:opt_theta}
	&\begin{array}{cl} \underset{K~{\rm stabilizing}}{\inf}&  \left\|\mathcal{F}(P;K)) \right\|\\
	\text{ s.t. }& K \text{ spatially-invariant}\\
	&\Phi^x, \Phi^u \text{ have band size }M \end{array}\\
	= &~
	\begin{array}{cl} \underset{\theta \in \oRs}{\inf}&  \left\| \left[ \begin{array}{cc} C_1 & D_{12} \end{array} \right] \left[ \begin{array}{c} F \theta + L \\ \chi \theta  + \eta \end{array} \right] B_1\right\|\\
	\text{ s.t. }& \theta \text{ spatially-invariant with band size }M
	\end{array} 
	\end{aligned}\end{equation}
	where $F, L, \chi, \eta$ are defined as in Theorem~\ref{thm:infDim}.
In the case of an $\h_2$ norm objective, the constrained controller design problem \eqref{eq:opt_theta} reduces further to a standard model matching problem \cite{francis1987course} with $(2M + 1)$ transfer function parameters.

\subsection{Optimal $\h_2$ Design Problem}  \label{subsec:modelmatch}
In examples throughout this paper, we measure closed-loop performance with the $\h_2$ norm.
One physical interpretation of this is that $\{ w_n\}$ is modeled as a mutually uncorrelated white stochastic process and the steady-state variance of the fluctuation of output $z$ then provides a measure of performance that can be calculated as the closed-loop $\h_2$ norm \cite{dullerud2013course}. The optimal controller design problem in this setting is 
	\begin{equation} \label{eq:ClosedLoopH2}
	 \inf_K~ \left\|\mathcal{F}(P;K) \right\|_{\h_2}^2 = ~\inf_K ~\Big\| \left[ \begin{array}{cc} C_1 & D_{12} \end{array} \right] \left[ \begin{array}{c}\Phi^x \\ \Phi^u \end{array} \right] B_1\Big\|_{\h_2}^2
	 \end{equation}

In the case of a spatially-invariant plant $P$, the stabilizing controller $K$ that optimizes \eqref{eq:ClosedLoopH2} is known to be a static (in time) spatial convolution kernel that can be solved for analytically using standard algebraic Riccati equation techniques in the finite space setting and using the techniques of \cite{bamieh2002distributed} in the infinite space setting. With the added closed-loop band size constraints however, the solution and its structural properties remain an open problem.

The structure of circulant and Toeplitz transfer matrices simplify the computation of the $\h_2$ norm in the spatially-invariant setting. Given a spatially-invariant system $H \in \oRs$, the $\h_2$ norm of the $n^{\rm th}$ row of the circulant matrix representation \eqref{eq:Kcirculant} or infinite Toeplitz matrix representation \eqref{eq:KToeplitz} of $H$ corresponds to the $L^2$ norm of the output at spatial site $n$ subject to impulse disturbances at all inputs. Similarly, the $\h_2$ norm of the $n^{\rm th}$ column corresponds to the $L^2$ norm of the output (at all spatial sites) subject to an impulse disturbance at spatial site $n$. The $\h_2$ norm of any row or any column of \eqref{eq:Kcirculant} or \eqref{eq:KToeplitz} will be equivalent and can be computed as:
	\begin{itemize}
	\item Finite Space Setting ($\G = \Z_N$):
		\begin{equation} \begin{aligned} \label{eq:H2normCirculant}
		 \sum_{m=0}^{N-1} \| H_m\|_{\h_2}^2 = \| H e_j \|_{\h_2}^2
		  = \frac{1}{N^2} \| H \|_{\h^2}^2,
		 \end{aligned}  \end{equation}
		 where $H e_j$ denotes the product of the circulant matrix $H$ with the $j^{\text{th}}$ standard basis vector  $e_j \in \R^N$ (e.g. $e_1 := \left[ \begin{array}{cccc} 1 & 0 & \cdots & 0 \end{array} \right]^T$).
		
	\item Infinite Space Setting ($\G = \Z$), with finite band size $M$:
		\begin{equation} \begin{aligned}  \label{eq:H2normToeplitz}
		  \sum_{m=-M}^{M} \| H_m\|_{\h_2}^2 
		  = \| H  e_j \|_{\h_2}^2 
		  = \frac{1}{(2M+1)^2} \| H\|_{\h_2}^2, 
		\end{aligned}  \end{equation}
		 where $H  e_j$ can be viewed as the product of the infinite-dimensional Toeplitz matrix $H$ with the infinite array $e_j =\left[ \begin{array}{ccccc} \cdots e_j(j-1) & e_j(j) & e_j(j+1) & \cdots \end{array} \right]^T =\left[ \begin{array}{ccccc} \cdots 0 & 1 & 0 & \cdots \end{array} \right]^T$; formally $H e_j$ is the convolution of 
$\{H_n\}_{n \in \Z}$ with the sequence defined by $e_j$.
	\end{itemize}
Note that the choice of index $j$ in both \eqref{eq:H2normCirculant} and \eqref{eq:H2normToeplitz} is arbitrary due to spatial invariance.
We refer to \eqref{eq:H2normCirculant} and \eqref{eq:H2normToeplitz} as the $\h_2$ norm \emph{per spatial site} of a spatially-invariant system. 
Thus in the spatially-invariant setting, optimizing the $\h_2$ norm of the system \eqref{eq:ClosedLoopH2} is equivalent to optimizing the $\h_2$ norm \emph{per spatial site}:

	\begin{equation}  \label{eq:h2objective}
		\left\| \left[ \begin{array}{cc} C_1 & D_{12} \end{array} \right] \left[ \begin{array}{c}\Phi^x \\ \Phi^u \end{array} \right] B_1 e_j\right\|_{\h_2}^2 
	\end{equation}
Employing our explicit closed-loop parameterizations, the closed-loop constrained controller design problem with $\h_2$ norm (per spatial cite) objective \eqref{eq:h2objective} reduces as follows:

	\begin{subequations} \begin{align} \label{eq:almostModelMatching}
	&\begin{array}{cc} 
		\underset{\theta \in \oRs}{\inf}&  \left\| \left[ \begin{array}{cc} C_1 & D_{12} \end{array} \right] \left[ \begin{array}{c} F \theta + L \\ \chi \theta  + \eta \end{array} \right] B_1 e_j\right\|_{\h_2}^2\\
	\text{ s.t. }&\theta \text{ spatially-invariant with band size } M
	\end{array} \\
	=&\begin{array}{cl}
		\underset{\theta \in \oRs}{\inf}& \left\| \left(C_1 L + D_{12} \eta \right) B_1 + \left( C_1 F + D_{12} \chi \right)~ \theta~ B_1e_j \right\|_{\h_2}^2\\
	\text{ s.t. }& \theta \text{ spatially-invariant with band size } M
	\end{array} \\
	\overset{(1)}{=}&\begin{array}{cc}
	 \underset{\vartheta \in \oRs}{\inf} & \| H + U \vartheta V \|_{\h_2}^2,
	\end{array} \label{eq:modelMatching}
	\end{align}
	\end{subequations}
where $F,~L$ and $\eta$ are defined in Theorems \ref{thm:nthOrder} and \ref{thm:infDim}, $H$ is composed of the nonzero entries of $\left(C_1 L + D_{12} \eta \right) B_1 e_j$, and $U, ~V$ are constructed accordingly from $ \left( C_1 F + D_{12} \chi \right)$ and $B_1 e_j$. The variable $\vartheta$ includes all non-zero components of the convolution kernel that specifies the parameter $\theta$, i.e.
	$$
		\vartheta(s) := \left[ \begin{array}{c} \theta_{-M}(s) \\ \vdots \\ \theta_M(s) \end{array} \right].
	$$
Thus
even in the infinite space setting, this problem has \emph{finitely many} transfer function parameters, 
provided a constraint of closed-loop band size $M < \infty$ is imposed.
For equality (1) to hold, we assume that $C_1$ and $D_{12}$ have finite band size so that $\left(C_1 L + D_{12} \eta \right) B_1 e_j$ has only finitely many nonzero entries. Examples of this procedure are provided in Section \ref{sec:applications}.

We refer to the form  
\eqref{eq:modelMatching} as a \emph{standard model matching problem} and refer back to this form throughout this paper. The optimal solution of this model matching problem can be computed analytically using the techniques of \cite[Ch. 6]{francis1987course}, when an inner-outer factorization $U = U_i U_o$ is available\footnote{
Such a factorization exists whenever $U$ is stable and strictly proper \cite{francis1987course}; it can be shown that this is always the case for $U$ constructed from our parameterization procedure
}. Even when such a factorization can not be computed, this problem is still tractable to solve numerically.

\section{Closed-Loop Parameterizations for Spatially-Varying Systems} \label{sec:spatiallyVarying}

In this section we present an explicit parameterization of all achievable stabilized closed-loop mappings for \emph{spatially-varying systems} with decoupled subsystem dynamics, generalizing the results of Sections \ref{sec:firstOrder} and \ref{sec:backstepping}.

\subsection{Locally $\first$ Order Systems}
We begin by analyzing the case of first-order subsystems:
\be
   \label{eq:firstOrderFinite}
	\dot{x}_n ~=~ a_n x_n~+~ w_n ~+~ b_nu_n, ~~n \in \Z_N, 
\ee
We assume \eqref{eq:firstOrderFinite} is controllable so that $b_n \ne 0$ for each $n$. We refer to systems of this form as {\em locally 1\textsuperscript{st} order}. This is a generalization of the class of {spatially-invariant locally 1\textsuperscript{st} order} systems presented in Section \ref{sec:1stOrder}.

At each spatial site $n$ the local state $x_n$ is scalar-valued, but the overall state $x$ is a vector of dimension $N$ composed of all the local states. 
In vector form, \eqref{eq:firstOrderFinite} is written as
\[
	\dot{x}  ~=~ A x~+~ w ~+~ B_2u , 
\]
where the ``A-matrix'' and ``B-matrix", $A := {\rm diag}\{a_n\}_{n \in \Z_N}$ and $B_2 := {\rm diag} \{b_n\}_{n \in \Z_N} $, are finite-dimensional diagonal matrices. For $A$ of this form, 
	\begin{equation} \begin{aligned} \label{eq1}
	\left( sI  - A \right)^{- 1}  = \left( sI - {\rm diag}  \{a_n\}\right)^{-1} = {\rm diag} \left\{ \frac{1}{s-a_n}  \right\},
	\end{aligned} \end{equation}
so that the affine relation \eqref{eq:newAffineConstraint} (with $p=0$) is simple to state in terms of the rows of the relevant matrices as
\begin{equation} \label{eq:rowConstraint}
	{\rm row}_n 	\!\lb	 \Phi^x 	\rb 
		 = \frac{1}{s-a_n}  ~{\rm row}_n\!\lb I + B_2 \Phi^u \rb .
\end{equation}
Using \eqref{eq:rowConstraint}, we derive an explicit parameterization of all achievable stabilized closed-loop mappings for the locally $\first$ order system \eqref{eq:firstOrderFinite}, as stated in the following theorem.

\begin{thm} \label{thm:finiteFirstOrder}
A controller $u = Kx$ is stabilizing for the locally $\first$ order system \eqref{eq:firstOrderFinite}
if and only if the resulting closed-loop mappings are of the form 
	\begin{eqnarray}  \label{eq:phiXloc1stOrder}
	\Phi^x(s) &=& B_2 \cdot  {\rm diag}\{\gamma_n\} \theta(s) + {\rm diag}\{\gamma_n\}\\
	\Phi^u(s) &=& {\rm diag} \{\alpha_n\} \theta(s) + {\rm diag}\{\beta_n\}, \label{eq:phiUloc1stOrder}
	\end{eqnarray}
for some $\theta \in \oRs$, where 
$\alpha_n, \gamma_n,$ and $\beta_n$ are defined as follows\footnote{
The choice of pole at $-p = -1$ in the definitions of $\alpha_n, \beta_n, \gamma_n$ arbitrary, analogous formulations for other stable poles choices could also be derived.}:
	$$\begin{cases}
	\alpha_n:= 1,~ \beta_n := 0,~ \gamma_n := \frac{1}{s - a_n},~~~~~ \text{if } {\rm Re}(a_n) < 0,\\
	\alpha_n:= \frac{s-a_n}{b_n(s+1)},~ \beta_n := - \frac{a_n+1}{b_n(s+1)},~ \gamma_n := \frac{1}{s+1},~~ \text{else.}
	 \end{cases} $$
\end{thm}

\begin{IEEEproof} See Appendix~\ref{app:proofs}. \end{IEEEproof}

We next extend these spatially-varying results to the case of higher order subsystems.

\subsection{Locally finite-dimensional Subsystems}
We next consider a distributed system with dynamics of the form
	\begin{equation}\label{eq:nthOrderDist}
	\dot{x}_n = A^{(n)} x_n + B^{(n)}_1 w_n + B^{(n)}_2 u_n, ~ n \in \Z_N,
	\end{equation}
with each $(A^{(n)} + I, B^{(n)}_2)$ of the form 
	\begin{equation*} \begin{aligned}
		({A^{(n)}}+I)&= \left[ \begin{array}{cccccc} -a_1^{(n)} I & -a_2^{(n)} I  & \cdots & -a_{r_n}^{(n)} I\\
						      I & 0 &  \cdots & 0 \\ 
						      & \ddots & \ddots  \\
						       & &  & 0 
				       \end{array} \right],\\
		{B}^{(n)}_2&= \left[ \begin{array}{ccccc} I & 0 & 0 & \cdots & 0\end{array} \right]^T
	\end{aligned} \end{equation*}
and each local state of the form
	$$
		x_n = \left[ \begin{array}{cccc} x_{n1}^T & x_{n2}^T & \cdots & x_{nr_n}^T \end{array} \right]^T
	$$
where $r_n$ denotes the order of subsystem $n$. We refer to systems of the form \eqref{eq:nthOrderDist} as \emph{locally finite-dimensional}. 
In vector form \eqref{eq:nthOrderDist} is written as
	$$
		 \dot{x} = A x + B_1 w + B_2 u,
	$$
where $A = { \rm diag} \{A^{(n)}\}_{n \in \Z_N}, B_2 =  { \rm diag} \{B^{(n)}_2\}_{n \in \Z_N}.$  A parameterization of all achievable stabilized closed-loop mappings for this class of systems is stated in the following theorem, whose proof is provided in Appendix~\ref{app:proofs}.
	\begin{thm} \label{thm:nthOrderDist} Let $\Phi^x_{ij}$ and $\Phi^u_{ij}$ denote the $(i,j)$ entries of the closed-loop mappings $\Phi^x$ and $\Phi^u$ respectively. Then the corresponding controller $u = Kx$ is stabilizing for the locally finite-dimensional system \eqref{eq:nthOrderDist} if and only if  $\Phi^x_{ij}$ and $\Phi^u_{ij}$ are of the form
		\begin{equation} \begin{aligned} \label{eq:nthOrderDist_phi}
		\Phi^x_{ij}(s) &=  \begin{cases}
		F(s) \theta_{ij}(s) + L(s),~ i = j\\
		F(s) \theta_{ij}(s),~~ i \ne j
		\end{cases}\\
		\Phi^u_{ij}(s) &=  \begin{cases}
		\chi^i(s) \theta_{ij}(s) + \eta^i(s),~ i = j\\
		\chi^i(s) \theta_{ij}(s),~~ i \ne j,
		\end{cases}
		\end{aligned} \end{equation}
	for some $\{\theta_{ij}\} \in \oRs$,
	where $F(s)$ and $L(s)$ are defined in \eqref{eq:Phi_nthOrder}, $\chi_i(s):= 1 +\frac{a_1^i}{s+1} +  \cdots + \frac{a_{r_i}^i}{(s+1)^{r_i}},$ and $\eta^i:= \left[ \begin{array}{ccc} \eta_1^i(s) & \cdots & \eta_{r_i}^i(s) \end{array} \right]$ with $\eta^i_k:=\sum_{\ell=k}^{r_i} \frac{a^i_{\ell}}{(s+1)^{\ell+ 1-k}}I$.
	\end{thm}
%

%%%%%%%%%%%%%%%%%%%%%%%% 

\section{Applications} \label{sec:applications}
\subsection{Consensus of first-order Subsystems} \label{sec:1stOrder}
We apply the parameterizations of Section \ref{sec:firstOrder}
to derive analytic solutions to the optimal closed-loop structured $\h_2$ design problem for consensus of first-order subsystems \eqref{eq:1storder} with performance output 
$$z = \left[ \begin{array}{c} y \\ \gamma u \end{array} \right] = \left[ \begin{array}{c} Cx \\ \gamma u \end{array} \right]  =  \left[ \begin{array}{cc} C & 0 \\ 0 & \gamma I \end{array} \right] \left[ \begin{array}{c} \Phi^x \\ \Phi^u \end{array} \right]w, $$
where $C$ is a spatial convolution operator that captures some measure of consensus. 
In the finite space setting ($\G = \Z_N$) we restrict our choice of band size constraint to $M< \frac{N}{2}$ so that the constraint is nontrivial.

By Theorem \ref{thm:infFirstOrder}, a spatially-invariant controller $u=Kx$ is stabilizing for \eqref{eq:1storder} if and only if the closed-loops satisfy
	\begin{equation} \begin{aligned} \label{eq:PhiFromThetaInf}
	\Phi^x &= (sI+I)^{-1} \theta + (sI+I)^{-1}, \\
	\Phi^u & = s(sI+I)^{-1}\theta - (sI+I)^{-1}
	\end{aligned} \end{equation}
	for some spatially-invariant system $\theta = \{ \theta_n \}_{ n \in \cG} \in \oRs$. The closed-loop structured $\h_2$ design problem \eqref{eq:almostModelMatching} for this system may be written as:
	\begin{equation} \begin{aligned}   \label{eq:optFirstOrderFinite}
	J:=& \inf_{\theta \in \oRs}~ \left\| \left[ \begin{array}{c} \frac{1}{s+1} Ce_j \\ \frac{-\gamma}{s+1} e_j \end{array} \right] + \left[ \begin{array}{c} \frac{1}{s+1} C \\ \frac{\gamma s } {s+1} I \end{array} \right] \theta e_j \right\|_{\h_2}^2 \\
	& ~~\text{s.t.}~~~ \theta \text{ spatially-invariant with band size }M\\
	\end{aligned} \end{equation}
	\begin{rem} A problem of interest for consensus applications is to restrict the controller to only have access to \emph{relative} state measurements, e.g. measurements of the form $x_i - x_j$. However, when finite band size constraints are imposed on the closed loops there does not exist a relative feedback controller that achieves a finite solution to \eqref{eq:optFirstOrderFinite} for any nontrivial band size constraint \cite{jensen2020Gap}. Thus the relative feedback controller design problem is not addressed in this paper. 
	\end{rem}

As shown in Section \ref{subsec:modelmatch}, \eqref{eq:optFirstOrderFinite} can be converted to a standard unconstrained model matching problem  \eqref{eq:modelMatching} with $2M+1$ transfer function parameters.
We consider the specific cases of $M= 1,2$ to illustrate this procedure and highlight the usefulness of our parameterization in leading to analytic solutions.
For $M=1$, \eqref{eq:optFirstOrderFinite} is written as
	\begin{equation}\begin{aligned} \label{eq:optSparsity1}
		\underset{\theta_{-1}, \theta_0, \theta_1 \in \oRs}{\inf}  \left\|  \left[ \begin{array}{c} \frac{1}{s+1} Ce_1 \\ \frac{-\gamma}{s+1} e_1 \end{array} \right]	  + {U} \left[ \begin{array}{c} \theta_{-1} \\ \theta_1 \\ \theta_{0} \end{array} \right] \right\|_{\h_2}^2,
		\end{aligned} \end{equation}
 where ${U}:= \left[ \begin{array}{ccc} {u}_{-1} & {u}_0 &{u}_1 \end{array} \right]$ with ${u}_{-1}, {u}_0, {u}_1$ the last, second and first column of $\left[ \begin{array}{c} \frac{1}{s+1} C \\ \frac{\gamma s } {s+1} I \end{array} \right]$ respectively, i.e. the columns which correspond to the nonzero entries of $\{\theta_n\}$. 
A similar formulation holds for $M = 2$. We use these formulations to compute analytic solutions for the optimal closed-loop norm with closed-loop band size constraints of $M=1,2$ for the following two measures of consensus:
\begin{enumerate}
\item \textbf{Local Error}
	$$y_n := x_n - x_{n-1}.$$
	Compactly we write $y = C^{\rm{LE}}x$, where $C^{\rm{LE}}$ is the spatial convolution operator with convolution kernel $c$ defined by $c_0 =1$, $c_{1} = -1$ and $c_k = 0$ for all $k \not \in \{0, 1\}$. $C^{\rm{LE}}$ can be represented in the finite space setting by the circulant matrix
	\begin{equation*} \label{eq:localErrorC}
	C^{\rm{LE}}  := \left[ \begin{array}{ccccc} 1 & 0 & \cdots & 0& -1\\ -1 &1 & \cdots &0&  0  \\ & \ddots & \ddots & \\ 0 &0 & \cdots & -1 & 1\end{array} \right],
	\end{equation*}
\item \textbf{Deviation from Average}
	$$y_n:= x_n - \frac{1}{N}\sum_{m = 0}^{N-1} x_m.$$
	Compactly we write $y = C^{\rm{Ave}}x$. In the finite space setting, $C^{\rm{Ave}}$ is given by the circulant matrix 
	\begin{equation*} \label{eq:localErrorC}
	C^{\rm{Ave}} := \frac{1}{N} \cdot \left[ \begin{array}{cccc} N-1 & -1 & \cdots & -1\\ -1 &N-1 & \cdots & -1  \\ & & \ddots & \\ -1 &-1 & \cdots  & N-1\end{array} \right].
	\end{equation*}
\end{enumerate}
These solutions are summarized in Table \ref{tab:firstOrder}. Analytic expressions for the optimal solution $\theta^{\rm opt}$ of \eqref{eq:optFirstOrderFinite} for these cases and detailed computations are provided in Appendix~\ref{app:consensus_computations}.

\begin{table}[h!]
  \begin{center}
    \caption{}
    \label{tab:firstOrder}
    \begin{tabular}{l|r} % <-- Alignments: 1st column left, 2nd middle and 3rd right, with vertical lines in between
      \textbf{} & \textbf{Optimal Closed-Loop Norm } \\
      & \textbf{ (per spatial site)}\\
      \hline
      \textbf{Local Error} &~\\
      ~ Sparsity $M = 1$ & $\frac{\gamma}{4} \left( (2 - \sqrt{2})^{1/2} + (2 + \sqrt{2})^{1/2} \right)$\\
      ~&$ \approx 0.653 \cdot \gamma$ \\
      ~ Sparsity $M = 2$ &$\frac{\gamma}{6}  \left(  (2 - \sqrt{3})^{1/2} + \sqrt{2} +   (2 + \sqrt{3})^{1/2}  \right)$\\
      ~&$ \approx 0.644 \cdot  \gamma$\\
      \textbf{Deviation From Ave.} &~\\
      ~ Sparsity $M = 1$ &$\gamma \left(\frac{1}{3} + \frac{1}{6}\sqrt{1 - \frac{3}{N}} \right)$\\
      ~ Sparsity $M = 2$ &$ \gamma \left( \frac{1}{8} + \frac{1}{10} \sqrt{1 - \frac{5}{N}}\right) $
    \end{tabular}
  \end{center}
\end{table}

To begin to physically interpret these solutions, we focus on the dependence of entries in Table \ref{tab:firstOrder} on system parameters $\gamma$ and $N$, rather than focusing on exact values. All entries scale linearly with control cost $\gamma$. However,
due to the added stochastic noise $\{ w_n\}$  the scaling in network size $N$ of the solution to \eqref{eq:optFirstOrderFinite} will differ based on the choice of consensus metric: for $C = C^{\rm{LE}}$ (first two rows of Table~\ref{tab:firstOrder}), the optimal cost is independent of $N$ and holds even for infinitely many subsystems; in contrast for $C = C^{\rm{Ave}}$ (last two rows of Table~\ref{tab:firstOrder}), the optimal cost\footnote{
We conjecture that the limit of the expressions for optimal cost provided in these last two rows as $N \rightarrow \infty$ provides the closed-loop norm per spatial site for the infinite space setting for $C^{\rm Ave}$ but do not formalize this here. } 
 is affine in $\sqrt{\frac{N-\nu}{N}}$ for constant $\nu$. Such dependencies on system parameters become clear with these analytic formulas, demonstrating a benefit of our proposed parameterization.

\subsection{Vehicular Platoons} \label{sec:2ndOrder}
We apply the parameterization presented in Theorem \ref{thm:infDim} to the vehicle platoons problem \eqref{eq:2ndorder}. The plant dynamics and performance output $z$ are written compactly as
	\begin{equation} \begin{aligned} \label{eq:vehicleVectorForm}
	\dot{x} &~=~ A x ~+~ B_1 w ~+~ B_2 u\\
	z & ~=~ \left[ \begin{array}{c} C \\ 0 \end{array} \right] x ~+~ \left[ \begin{array}{c} 0 \\ \gamma I \end{array} \right] u 
	\end{aligned} \end{equation}
where $A, B_1, B_2$ are the pointwise multiplication operators defined in Example~\ref{exmp:vehicleDynamics}. By Theorem \ref{thm:infDim}, a spatially-invariant controller $K$ is stabilizing for \eqref{eq:vehicleVectorForm} if and only if the corresponding closed-loop mappings $\Phi^x = \{ \Phi^x_n \}_{n \in \G}$ and $\Phi^u = \{ \Phi^u_n \}_{n \in \G}$ are of the form:
	\begin{equation} \begin{aligned} \label{eq:vehicleConsensusPhi}
	\Phi^x_n(s)&= \begin{cases}F \theta_n + L,~ n = 0\\ F \theta_n,~~~~~~ n \ne 0
	\end{cases}\\
	\Phi^u_n(s)&= \begin{cases} \chi \theta_n + \left[ \begin{array}{cc} \eta_1 & \eta_2 \end{array} \right], ~ n = 0\\ 
	 \chi \theta_n,~~~~~~~~~~~~~~~~~~~ n \ne 0
	\end{cases}
	\end{aligned} \end{equation}
 with $F(s) =  \left[ \begin{array}{c} \frac{1}{s+1} \\ \frac{1}{(s+1)^2} \end{array} \right]$, $L(s) = \left[ \begin{array}{cc} \frac{1}{s+1} & 0 \\ \frac{1}{(s+1)^2} & \frac{1}{s+1} \end{array} \right]$, $\chi(s) = \frac{s^2}{(s+1)^2}$,  $\eta_1(s) = \frac{-2s-1}{(s+1)^2}$ and 
 $$\theta_n =  \left[ \begin{array}{cc} \theta^{(1)}_n & \theta^{(2)}_n\end{array} \right] \in \oRs$$
a $1 \times 2$ transfer matrix parameter. 
 Denoting the spatially-invariant systems by $\theta^{(1)} = \{ \theta^{(1)}_n \}_{n \in \cG}, ~ \theta^{(2)} = \{ \theta^{(2)}_n \}_{n \in \cG}$,
the optimal $\h_2$  design problem \eqref{eq:almostModelMatching} for \eqref{eq:vehicleVectorForm} subject to closed-loop band size constraints can be written in terms of parameters $\theta^{(1)}$ and $\theta^{(2)}$ as
	\begin{equation} \begin{aligned} \label{eq:optVehicleConsensus}
	&\inf_{\theta^{(1)}, \theta^{(2)} \in \oRs} ~ \left\| \left[ \begin{array}{cc} C & 0 \\ 0 & \gamma I \end{array} \right]
	\left[ \begin{array}{c} F \left[ \begin{array}{cc}\theta^{(1)} &\theta^{(2)}\end{array} \right]  + L \\ \chi \left[ \begin{array}{cc}\theta^{(1)} &\theta^{(2)} \end{array} \right] + \eta \end{array} \right] B_1 e_j\right\|_{\h_2}^2 \\
	&~~~~~\text{s.t.} ~~~~~ \theta^{(1)} ,~\theta^{(2)} \text{ spatially-invariant with band size } M\\
	&\overset{(1)}{=}~   \inf_{\theta^{(1)} \in \oRs} ~ \left\| \left[ \begin{array}{cc} C & 0 \\ 0 & \gamma I \end{array} \right]
	\left[ \begin{array}{c} F \theta_1 + F \\ \chi \theta^{(1)} + \eta_1 \end{array} \right] e_j \right\|_{\h_2}^2\\
	&~~~~~~~\text{s.t.} ~~~~~ \theta^{(1)} \text{ spatially-invariant with band size } M\\
	&=~   \inf_{\theta^{(1)} \in \oRs} ~ \left\| \left[ \begin{array}{c} CF \\ \gamma \eta_1 \end{array} \right] e_j + \left[ \begin{array}{c} CF \\ \gamma \chi I\end{array} \right] \theta^{(1)} e_j\right\|_{\h_2}^2\\
	&~~~~~~~\text{s.t.} ~~~~~ \theta^{(1)} \text{ spatially-invariant with band size } M
	\end{aligned} \end{equation}
where $F, L, \eta_1$ represent pointwise multiplication by the finite-dimensional transfer matrices $F(s), L(s), \eta_1(s)$. The equality $\overset{(1)}{=}$ follows from the fact that $B_1 = \left[ \begin{array}{c} I \\ 0 \end{array} \right]$.
 We consider $C$ corresponding to one of the following measures of consensus:
 \begin{itemize}
 \item \textbf{Local error of vehicle position:}
 	$$y_n := (C^{\rm LE}x)_n =  \left[ \begin{array}{cc} 0 & 1 \end{array} \right] (x_n - x_{n-1}).$$
\item \textbf{Deviation from average of vehicle position:}
	$$y_n := (C^{\rm Ave}x)_n =  \left[ \begin{array}{cc} 0 & 1 \end{array} \right] x_n - \frac{1}{N} \sum_{i=1}^N \left[ \begin{array}{cc} 0 & 1 \end{array} \right] x_i,$$
	with $N$ the number of subsystems. 
\end{itemize}

\begin{figure} [h!]
     \centering
         \includegraphics[width=94mm]{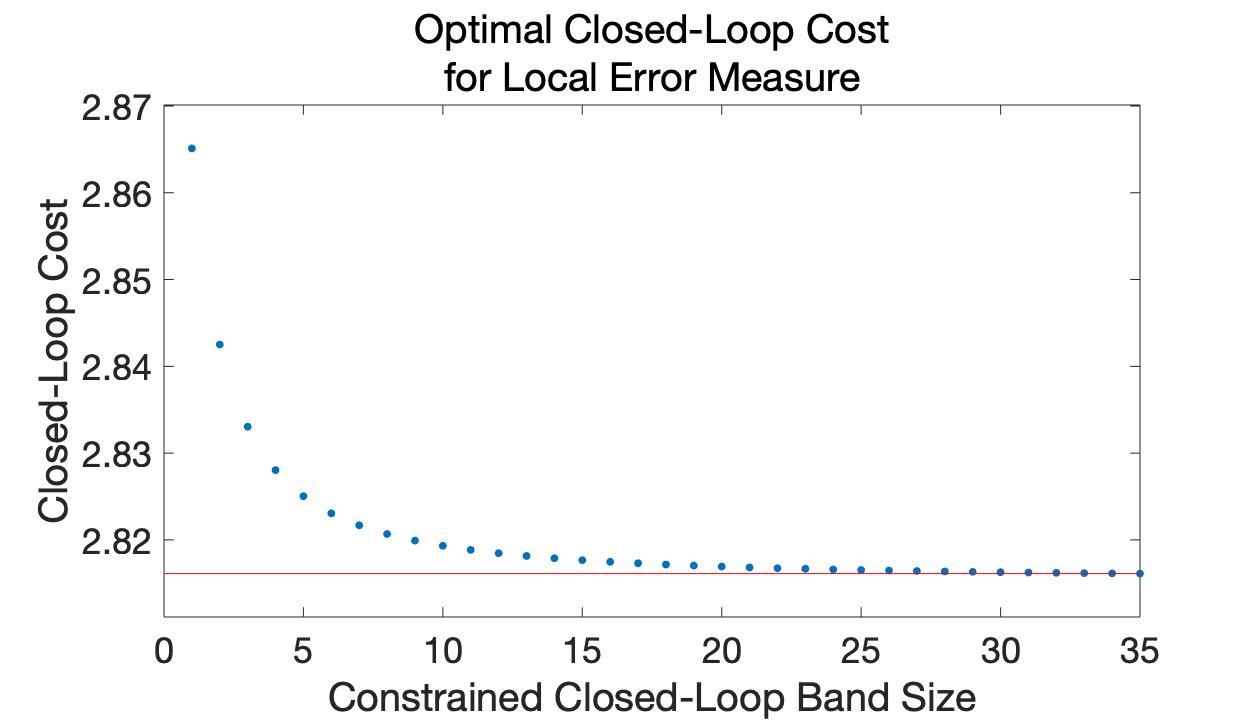}
         
         	         \vspace{.4cm}

         \includegraphics[width=94mm]{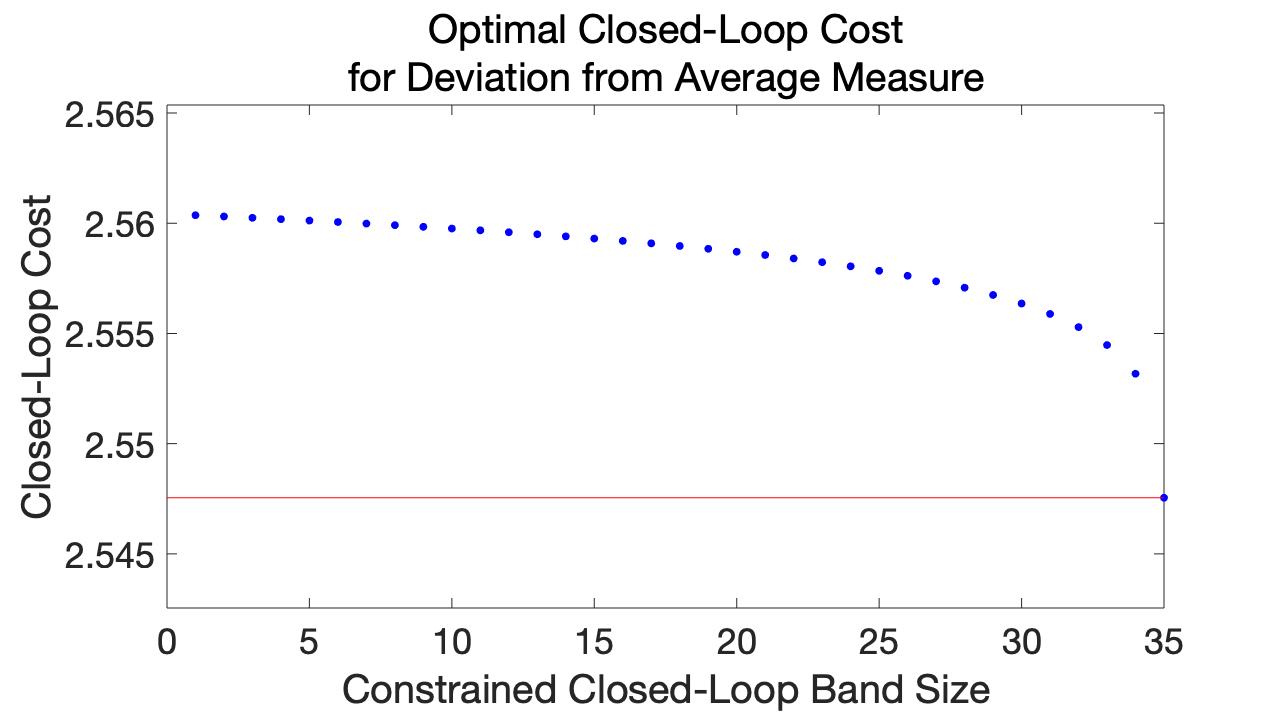}
         \caption{The optimal closed-loop cost $(\h_2$ norm) of the constrained controller design problem \eqref{eq:optVehicleConsensus} for the vehicle platoons problem with local error metric (top) and deviation from average metric (bottom) is plotted against the spatial sparsity extent constraint $M$ imposed on the closed-loop mappings for $N = 71$ vehicles and control cost weighting $\gamma = 3$. The red line illustrates the optimal for the unconstrained problem, i.e. for band size $M = \frac{N-1}{2}= 35$. 
         }
         \label{fig:devFromAve}
\end{figure}

For either of these choices of $C$, we follow the procedure of Section \ref{subsec:modelmatch} to reduce \eqref{eq:optVehicleConsensus} to an unconstrained model-matching problem \eqref{eq:modelMatching} with 
$2M+1$ transfer function optimization variables $\vartheta:= \left[ \begin{array}{ccc} \theta_{1,-M} & \cdots & \theta_{1,M} \end{array} \right]^T$ and solve this problem numerically for various choices of band size constraint $M$. 
The results for the case of a control weight $\gamma = 3$ and $N = 71$ subsystems are illustrated in Figure~\ref{fig:devFromAve}: the optimal closed-loop cost is plotted as a function of closed-loop band size $M$ for a local error objective (top) and a deviation from average objective (bottom). 
As band size $M$ increases, the closed-loop maps have less constrained structure, and the corresponding closed-loop cost decreases toward the unconstrained optimal illustrated by the red lines.

In the local error case (top) the convergence appears roughly exponential and the improvement in closed-loop cost with each new communication link diminishes as the amount of communication increases, i.e.
 after a certain point the amount of performance gained for each new communication link becomes negligible. The convergence rate for the deviation from average measure (bottom) is qualitatively different; in this case the improvement in closed-loop cost with each new communication link increases as the amount of communication increases. Quantifying these decay rates and understanding the differences for these two measures is the subject of future work. It is known that the optimal unconstrained controller for the local error measure is a spatial convolution operator with kernel that decays exponentially \cite{bamieh2002distributed}. An interesting open question is how to formalize such a notion of decay rate for the \emph{dynamic} closed-loop mappings in this setting and prove whether such a structural property explains the apparent exponential decay in Figure 1 (top).

\section{Structured Controller Implementation} \label{sec:structured}

The closed-loop band size of $M$ constraint in \eqref{eq:optAnyNorm} ensures that a disturbance entering into any subsystem does not affect neighboring subsystems more than a distance of $M$ away \emph{for all time} - e.g. a band size of one means that disturbances at subsystem $n$ may only affect subsystems $n-1, n,$ and $n+1$ for all time. This disturbance {localization} property is useful in certain applications where it is desirable to completely eliminate disturbance effects outside some spatial region for all time. However, in many applications this property is overly aggressive to achieve desired performance, and the true benefit of this constraint is that the closed-loop band size carries over to the locality of an \emph{implementation} of the resulting controller. 

The importance of the structure of a state-space realization of a controller as opposed to the structure of the controller transfer matrix has only recently been emphasized in the literature, e.g. \cite{lessard2013structured,rantzer,vamsi2015optimal,lessard2015optimal}. The structure of state-space realizations of controllers that result in structured closed-loops, as 
derived from the SLS framework, has been analyzed in the discrete-time FIR setting in \cite{anderson2017structured} and in the IIR continuous time setting in \cite{jensen2020structured}. In this section, we examine the relation between closed-loop transfer function structure and corresponding controller realization structure with a case study.

We begin by introducing the following definition to formally analyze the structure of a \emph{realization} of a system. 

	\begin{defn} \label{def:structured}
	Let  
		\begin{equation} \label{eq:ss}
			G(s) = C(sI-A)^{-1}B +D
		\end{equation}
		with circulant matrices (i.e. spatial convolution operators) $A, B, C, D$
	be a state-space realization of a spatially-invariant system $G$ over $\Z_N$.
	This realization \eqref{eq:ss} is said to be a \emph{structured realization} with band size $M$ if the matrices $A, B, C, D$  each have band size $M$ (according to Definition~\ref{def:bandsize}). When such a realization exists, $G$ is said to be \emph{structured-realizable} with band size $M$.
	\end{defn}

	This notion of structured-realizability can be extended to spatially-invariant systems over a countably infinite spatial domain as follows. A spatially-invariant system $G$ with input $u$ and output $y$ is said to be \emph{structured-realizable} with band size $M$ if there exists an implementation for which the output at spatial location $n$, $y_n(t)$, can be computed using only inputs $u_m(\tau)$ with $|m-n| <M$ and $\tau \le t$. This idea can be formalized using the framework for infinite-dimensional systems presented in \cite{curtain2020introduction} although we omit these technical details here. 	
	
\begin{lem} \label{lem:implementation} 
Let $\theta^{\rm opt}$ denote the optimizer of \eqref{eq:optFirstOrderFinite} for $C = C^{\rm LE}$ with closed-loop band size constraint $M = 1$, and let 
$K^{\rm opt}$ denote the corresponding controller recovered from $\theta^{\rm opt}$, i.e. $K^{\rm opt}$ is the 
$\h_2$ optimal controller for consensus of first order subsystems under a local error metric, subject to closed-loop band size constraint $M = 1$. Then 
$K^{\rm opt}$ has a structured realization of band size $M = 1$, i.e. of the form: 
	\begin{equation} \begin{aligned} \label{eq:structuredRealization}
	 	\dot{\psi}_m &= \sum_{i = m-1}^{m+1}A^{(k)}_ i \psi_i + B^{(k)} x_i  \\
	 	u_m &= \sum_{i = m-1}^{m+1}C^{(k)}_ i \psi_i + D^{(k)} x_i, ~ m \in \cG
	\end{aligned} \end{equation} 
where $\psi_m$ denotes the state of subcontroller $m$. 
\end{lem}

\begin{figure}[t]
	 	\begin{centering}
 		\includegraphics[width=.45\textwidth]{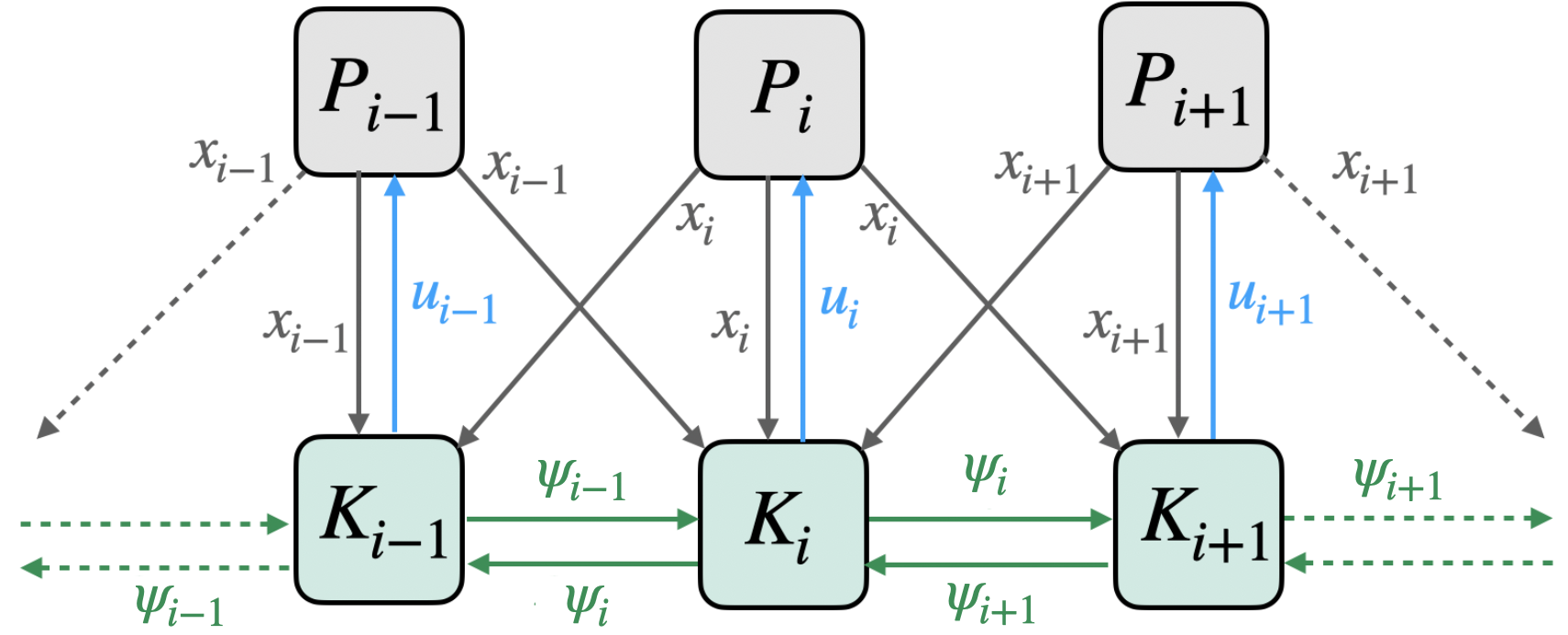}
 		 \caption{\footnotesize Implementation of optimal controller with spatial sparsity extent of one imposed on the closed-loop mappings. Each subplant $P_i$ sends its local state $x_i$ to subcontrollers $K_i$, $K_{i-1}$, and $K_{i+1}$.  Each subcontroller $K_i$ sends its local state $\psi_i: = \left[\xi_i^T ~\zeta_i^T \right] $ to neighboring subcontrollers $K_{i-1}$ and $K_{i+1}$. Each subcontroller $K_i$ provides the local control action $u_i$ to plant $P_i$ based on this information exchange.} \label{fig:controllerImplementation}
		 \end{centering}
	\end{figure}

The implementation \eqref{eq:structuredRealization} of $K^{\rm opt}$ is such that each local subcontroller state is computed using only the states of subcontrollers and subsystems within a neighborhood of size one, i.e. $\psi_m(t)$ is computed from $\psi_j$ and $x_j$ for $j = m-1, m$ and $m+1$. (see Figure \ref{fig:controllerImplementation}).  The controller $K^{\rm opt}$ itself
in general is \emph{not} a transfer matrix with finite band size. This highlights the difference between the structure of a controller {transfer function} (Definition~\ref{def:bandsize}) and the structure of a corresponding {realization} (Definition~\ref{def:structured}); this distinction has been recently emphasized in e.g. \cite{rantzer,vamsi2015optimal, lessard2013structured, jensen2020Gap}. 

\begin{IEEEproof}
We leverage (a slight modification of) the implementation suggested in \cite{wang2019system}, and implement the controller $u =K^{\rm opt}x$ as follows:
	\begin{equation}  \label{eq:controllerImplementation}
	\begin{array}{ll}
	v = (I - (s+p) \Phi^x) v + x &=:  \widetilde{\Phi}^x v + x  \\
	& =: \tilde{x} + x\\
	u = ((s+p) \Phi^u) v &=:\widetilde{\Phi}^u v, 
	\end{array} \end{equation}
for ${\rm Re}(p) > 0$ where $\Phi^x, \Phi^u$ are the closed-loop mappings resulting from the optimal controller $K^{\rm opt}$
(see Figure \ref{fig:controller}). Direct computations show that the mapping from $x$ to $u$ defined by this feedback diagram is given by $u = \Phi^u (\Phi^x)^{-1} x = K^{\rm opt} x$. To see that the feedback loop from $x$ to $v$ in Figure~\ref{fig:controller} is well-posed, note that the affine subspace constraint \eqref{eq:newAffineConstraint} shows that $I - (s+p)\Phi^x = (A+pI) \Phi^x - B_2 \Phi^u$ is strictly proper. It can also be shown that stability of $\Phi^x$ and $\Phi^u$ imply internal stability of the feedback interconnection of this block diagram representation of $K^{\rm opt}$ with plant $P$.  

	\begin{figure}[t]
	 	\begin{centering}
 		\includegraphics[width=.48\textwidth]{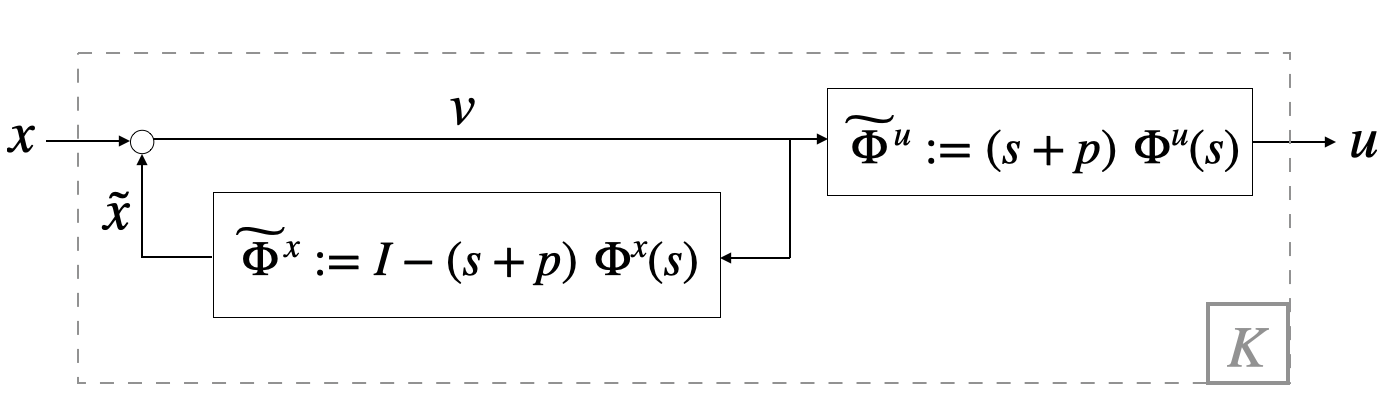}
 		 \caption{\footnotesize Implementation of controller $K = \Phi^u(\Phi^x)^{-1}$ via closed-loop mappings to preserve structure. Any choice of ${\rm Re}(p) >0$ will ensure internal stability of the closed-loop mappings for $\Phi^u$ and $\Phi^x$ stable.} \label{fig:controller}
		 \end{centering}
	\end{figure}
The spatially-invariant systems $\widetilde{\Phi}^x$ and $\widetilde{\Phi}^u$ (see Eq. \eqref{eq:controllerImplementation} and Figure~\ref{fig:controller}) have the same band size as $\Phi^x$ and $\Phi^u$ of $M = 1$. Thus $\widetilde{\Phi}^x$ and $\widetilde{\Phi}^u$ are specified by the 3 nonzero components of their convolution kernels. Specifically,
	\begin{equation} \label{eq:tildePhi1}
	\left[ \begin{array}{c} \tilde{x}_n \\ u_n \end{array} \right] = \left[ \begin{array}{ccc} \widetilde{\Phi}^x_{-1} & \widetilde{\Phi}^x_0 & \widetilde{\Phi}^x_1 \\ \widetilde{\Phi}^u_{-1} & \widetilde{\Phi}^u_0 & \widetilde{\Phi}^u_1 \end{array} \right] \left[ \begin{array}{c} v_{n-1} \\ v_n \\ v_{n+1} \end{array} \right],
	\end{equation}
where $\tilde{x}_n$ is the $n^{\rm th}$ component of the output of $\widetilde{\Phi}^x$ and
$u_n$ is the output of the subcontroller at site $n$.
An explicit expression for the transfer function \eqref{eq:tildePhi1} is calculated from the formulas for $\Phi^x$ and $\Phi^u$ that are stated in Eq. \eqref{eq:optPhiXPhiU1stOrder} of Appendix~\ref{app:consensus_computations}. Realizations of \eqref{eq:tildePhi1} that lead to the structured realization \eqref{eq:structuredRealization} of $K^{\rm opt}$ are provided in Appendix~\ref{app:implementation}.
\end{IEEEproof}

\begin{rem} The realization \eqref{eq:structuredRealization} is not relative, i.e. {absolute} measurements of subsystem states are required  to implement the controller in this way. Thus, this implementation does not provide a solution to the relative feedback control problem addressed in \cite{jensen2020Gap}.
\end{rem}

\section{Conclusion}
An \emph{explicit} parameterization of the set of all achievable stabilized closed-loop mappings for subclasses of spatially-distributed systems was derived. 
In contrast to the {implicit} parameterization introduced by SLS, our explicit parameterization eliminated the need for temporal FIR approximations, allowing the $\h_2$ design problem to be converted to a standard unconstrained model matching problem and admitting analytic IIR solutions. A relation of our parameterization to the classical Youla parameterization was illustrated in a specific case. 
We studied two applications (consensus of first-order subsystems and the vehicular platoons problem) to comment on performance scalings with system size and structural constraints. The consensus example was also used to demonstrate the structure of controller implementations resulting from our parameterizations. Future work includes extending the parameterizations provided in this paper to more general settings (e.g. coupled dynamics and output feedback), and formally analyzing convergence rates observed numerically.

\bibliographystyle{ieeetr}
\bibliography{ACC_bib_NEW}{}

% or


\appendices

\section{Coupled Subsystem Dynamics} \label{app:coupled}
The results presented in this Section are straightforward modifications/ generalizations of the results of \cite{jensen2018optimal}. Detailed proofs of these results are presented in \cite{jensen2020topics}.  

\begin{thm} \label{thm:coupled}
Consider the system \eqref{eq:spatiallyInvGeneral} in feedback with a controller $u(t) = (Kx)(t)$ and assume that either
	\begin{enumerate}  [label=\roman*)]
		\item $A$ and $B$ are finite-dimensional matrices and $K(s)$ is a transfer function with finitely many inputs/outputs, or
		\item $A$ and $B$ are spatial convolution operators with convolution kernels $\{A_n\}_{n \in \cG}, \{B_n\}_{n \in \cG}$ each having only finitely many nonzero components, and $K$ is a spatially-invariant system. 
	\end{enumerate}
Then the following hold:
	\begin{enumerate}[label=\alph*)]
	\item When \eqref{eq:spatiallyInvGeneral}  has stable open-loop dynamics, $K$ is internally stabilizing if and only if the closed-loop mapping $\Phi^u \in \oRs$ and $\Phi^x$ is of the form 
		\begin{equation} 
			\Phi^x = (sI - A)^{-1} ( I + B_2 \Phi^u).
		\end{equation}
	\item Let ${\rm Re}(p)>0$. When the operator $B_2$ is invertible, $K$ is internally stabilizing if and only if the corresponding closed-loop mappings are of the form
		\begin{equation} \begin{aligned}
			\Phi^x &= \frac{1}{s+p} (I + \theta)\\
			\Phi^u & = \frac{1}{s+p} B_2^{-1} \left( \left(sI - A \right) \theta - (A+pI)\right),
		\end{aligned} \end{equation}
	for some $\theta \in \oRs$.
	\end{enumerate}
\end{thm}
In either setting, coupling between subsystem dynamics is described by the `off-diagonal' block entries of $A$ and $B$. $\Phi^x, \Phi^u$ and $\theta$ are each finite-dimensional transfer functions in case (i) and spatially-invariant systems in case (ii).

\section{Proofs of Theorems} \label{app:proofs}

\subsection{Proof of Theorem \ref{thm:infFirstOrder}} 
Assume that $K$ is spatially-invariant and provides a stabilizing controller for \eqref{eq:1stSpatiallyInv} in the case that
$\cG = \Z$ and ${\rm Re}(a) \ge 0$. Then following \cite{jensen2018optimal}, the definitions \eqref{eq:phi_inf} 
of $\Phi^x = \{ \Phi^x_n(s)\}_{n \in \Z}$ and $\Phi^u= \{ \Phi^u_n(s)\}_{n \in \Z}$ show that
	$(sI - aI ) \Phi^x - \Phi^u = I.$
Equivalently
	$$
		(s-a) \Phi^x_n(s) -  \Phi^u_n(s) = \begin{cases} 0,~ n \in \Z\setminus\{0\} \\ 1, ~ n = 0.\end{cases}
	$$
Then by the arguments presented for the finite space setting, $\Phi^x_0,$ and $\Phi^u_0$ are of the form:
	\begin{equation} \begin{aligned}
		\Phi^x_0(s) &= \frac{1}{s+p} \theta_0(s) + \frac{1}{s+p}I,\\
		\Phi^u_0(s) &= \frac{s-a}{s+p} \theta_0(s) - \frac{a+p}{s+p} I.
	\end{aligned}\end{equation}
Similarly it can be shown that $\Phi^x_n(s) = \frac{1}{s+p} \theta_n(s),~ \Phi^u_n(s) = \frac{s-a}{s+p} \theta_n(s)$ for $n \ne 0$. From these formulations we see that the spatially-invariant systems $\Phi^x, \Phi^u$ are of the form \eqref{eq:infFirstOrderX}, \eqref{eq:infFirstOrderU}. 
Conversely if $\Phi^x, \Phi^u$ are of the form \eqref{eq:infFirstOrderX}, \eqref{eq:infFirstOrderU}, then $\Phi^x, \Phi^u \in \oRs$ so that $K$ is stabilizing. \hfill $\blacksquare$

\subsection{Proof of Theorem \ref{thm:nthOrder}}
 If $\Phi^x, \Phi^u$ are of the form \eqref{eq:Phi_nthOrder} for some $\theta \in \oRs$, then $\Phi^x, \Phi^u \in \oRs$ and direct computations show that these mappings satisfy \eqref{eq:newAffineConstraint}.
Conversely, assume $\Phi^x, \Phi^u \in \oRs$ satisfy \eqref{eq:newAffineConstraint}. We will demonstrate through a back-stepping like procedure that $\Phi^x, \Phi^u$ are of the form \eqref{eq:Phi_nthOrder}.  

\subsubsection{Backstepping Algorithm}
Partition $\Phi^x$ by block rows as
	 $$\Phi^x = \left[ \begin{array}{cccc} \Phi^x_{11} & \Phi^x_{12} & \cdots & \Phi^x_{1r} \\\Phi^x_{21} & \Phi^x_{22} & \cdots & \Phi^x_{2r} \\\vdots \\ \Phi^x_{r1} & \Phi^x_{r2} & \cdots & \Phi^x_{rr} \end{array} \right] =:
	 \left[ \begin{array}{c} \Phi^x_1 \\ \Phi^x_2 \\  \vdots \\ \Phi^x_r\end{array} \right]$$

If $(A+I)$ and $B_2$ are of the form \eqref{eq:nthOrder}, then
	\begin{equation*} \begin{aligned}
	&\left(\left(s+1\right)I-\left(A+I\right) \right) =\\
	&~~~~~\left[ \begin{array}{ccccc} (s+1+a_1)I & a_2 I & \cdots & a_{r-1}I & a_rI\\
	-I & (s+1)I & \cdots & 0 &0 \\
	0 & -I & \cdots & 0 & 0 \\
	\vdots & & \ddots \\
	0 & 0 & \cdots &-I & (s+1)I \end{array} \right],
	\end{aligned} \end{equation*}
and for any $k \ne 1$, the $k^{\text{th}}$ block row of the affine constraint \eqref{eq:newAffineConstraint} can be rearranged as:
	\begin{equation} \label{eq:phiXk}
		 \Phi^x_k = \frac{1}{(s+1)} \left( E_k + \Phi^x_{k-1} \right), 
	\end{equation}
where $E_k$ is defined to be the matrix whose $k^{\text{th}}$ block entry is the identity, and all other entries are zeros:
	$$
		E_k:=\left[ \begin{array}{ccccccc} 0 & \cdots & 0 & I & 0 & \cdots & 0 \end{array} \right].
	$$
Back-substituting, each block row, $\Phi^x_k$, can be written in terms of the first block row, $\Phi^x_1$, as:
	\begin{equation} \begin{aligned} \label{eq:Phixk}
	\Phi^x_k &=\frac{1}{(s+1)^{k-1}} \Phi^x_1 +  \frac{1}{s+1} E_k + \frac{1}{(s+1)^2} E_{k-1} \\   
	&~~~~~+ \frac{1}{(s+1)^{k-1}}E_2
	\end{aligned} \end{equation}
From \eqref{eq:Phixk} it follows that $\Phi^x \in \oRs$ whenever $\Phi^x_1 \in \oRs$. 
Rearranging the first block row, $\Phi^x_1$, of \eqref{eq:newAffineConstraint} shows that 
	\begin{equation}\label{eq:phiUphiXk}
	 \Phi^u = (s+1+a_1)\Phi^x_1 - E_1 - \sum_{k=2}^r\Phi_k^x 
	 \end{equation}
Substituting \eqref{eq:Phixk} into \eqref{eq:phiUphiXk} shows that:
	\begin{equation} \label{eq:phiUk}
	\Phi^u_k = \begin{cases} \alpha(s) \Phi^x_{1k}(s) +\sum_{i=k}^r\frac{a_i}{(s+1)^{i+1-k}}I,~ k \ne 1\\
	\alpha(s) \Phi^x_{11}(s) - I,~ k = 1 \end{cases}
	\end{equation}
where $\alpha(s):= (s+1) + a_1 + \frac{a_2}{s+1} + \cdots + \frac{a_r}{(s+1)^{r-1}},$ and we have partitioned 
$\Phi^u$ and each $\Phi^x_k$ by block columns as:
	\begin{equation*} \begin{aligned}
		\Phi^u &= \left[ \begin{array}{cccc}\Phi^u_1 & \Phi^u_2 \cdots & \Phi^u_r \end{array} \right]
		 \Phi^x_k& =  \left[ \begin{array}{cccc} \Phi^x_{k1} &\Phi^x_{k2} & \cdots & \Phi^x_{kr} \end{array} \right]
	\end{aligned} \end{equation*}
 From \eqref{eq:phiUk} we see that if $\Phi^x_{1k} \in \oRs$ then the following is a necessary and sufficient condition for $\Phi^u_k \in \oRs$:
	$$
		\begin{cases} (s+1) \Phi^x_{1k} \in \oRs,~~ k \ne 1,\\
		(s+1) \Phi^x_{11}(s) - I \in \oRs,~~ k = 1. \end{cases}
	$$ 
	Equivalently,
	\begin{equation} \label{eq:phiX1k}
	\Phi^x_{1k}(s) = \begin{cases} \frac{1}{s+1} \theta_k(s),~ k \ne 1\\
	\frac{1}{s+1} \theta_k(s) + \frac{1}{s+1} I,~ k = 1 \end{cases}
	\end{equation}
for some $\theta_k \in \oRs$. Substituting \eqref{eq:phiX1k} into \eqref{eq:phiUk} and \eqref{eq:Phixk} shows that $\Phi^u$ and $\Phi^x$ are of the form \eqref{eq:Phi_nthOrder}. \hfill $\blacksquare$

\subsection{Proof of Theorem \ref{thm:infDim}}
 A direct application of the results of \cite{jensen2018optimal} shows that the spatially-invariant system $K$ stabilizes \eqref{eq:infDimZ} if and only if the resulting closed-loops $\Phi^x, \Phi^u \in \oRs$ and satisfy
	\begin{equation} \label{eq:infDimAffine}
	\left( (sI+I) - (A+I) \right) \Phi^x - B_2 \Phi^u = I,
	\end{equation}
where $A$ and $B_2$ are pointwise multiplication operators. Recall that $\Phi^x$ and $\Phi^u$ will be spatially-invariant by Proposition~\ref{prop:spatiallyInvProperties}. 
 The affine constraint \eqref{eq:infDimAffine} can be written equivalently in terms of the convolution kernels of $\Phi^x$ and $\Phi^u$ as
	\begin{equation} \label{eq:affineConstraintPointwise}
	\left[ \begin{array}{cc} sI + I -\left(A+I\right) & -B_2 \end{array} \right] \left[ \begin{array}{c} \Phi^x_n \\ \Phi^u_n \end{array} \right] =
	\begin{cases} 0,~  n \in \cG\setminus \{0\}\\ I, ~ n = 0 \end{cases}
	\end{equation}
and $\Phi^x, \Phi^u \in \oRs$ if and only if each $\Phi^x_n, \Phi^u_n \in \oRs$. Note that the two cases in this constraint lead to the two parameterizations of \eqref{eq:nthOrderPhiComponents}. By Theorem \ref{thm:nthOrder}, each $\Phi^x_0$ and $\Phi^u_0$ satisfy \eqref{eq:affineConstraintPointwise} if and only if 
they are of the form \eqref{eq:nthOrderPhiComponents}. It is straightforward to modify this argument to show an equivalent result for $\Phi^x_n, \Phi^u_n$ for $n \ne 0$. Thus, condition \eqref{eq:affineConstraintPointwise} is equivalent to the formulation \eqref{eq:nthOrderPhiComponents}. \hfill $\blacksquare$

\subsection{Proof of Theorem \ref{thm:finiteFirstOrder}}
Let $\Phi^x, \Phi^u$ denote the closed-loop mappings corresponding to a controller $K$ for
 \eqref{eq:firstOrderFinite}. If $\Phi^x, \Phi^u$ are of the forms \eqref{eq:phiXloc1stOrder}-\eqref{eq:phiUloc1stOrder} for some $\theta \in \oRs$, then they are both elements of $\oRs$ and direct computations show that \eqref{eq:newAffineConstraint} holds. Conversely, assume that $K$ is stabilizing. Then $\Phi^x, \Phi^u \in \oRs$, and following \cite{wang2019system}, direct computations show that $\Phi^x$ and $\Phi^u$ satisfy \eqref{eq:newAffineConstraint} with $p = 0$:
	\begin{equation} \label{eq:affine}
(sI-A) \Phi^x(s) = I + B_2 \Phi^u(s),
	\end{equation}
where $A = {\rm diag}\{a_m\}$, $B_2 ={\rm diag}\{b_m\}$. 
Equivalently, 
	\begin{equation} \label{eq:phix_ij}
	\Phi^x_{ij}(s) = \begin{cases}
	\frac{b_i}{s-a_i} \Phi^u_{ij}(s),~ i \ne j\\
	\frac{1}{s-a_i} + \frac{b_i}{s-a_i} \Phi^u_{ij}(s),~ i = j
	 \end{cases}
	 \end{equation}
where $\Phi^x_{ij}$ and $\Phi^u_{ij}$ are the $(i,j)$ components of $\Phi^x$ and $\Phi^u$. If ${\rm Re}(a_i) \ge 0$, then a necessary condition for $\Phi^u$ to be stable is that $\Phi^u_{ij}$ does not have a pole at $a_i$ for any $j$. Then, since $\Phi^u_{ij}$ is strictly proper, it must be of the form:
	\begin{equation} \label{eq:phiu_ij}
	\Phi^u_{ij}(s) = \begin{cases}
	\frac{s-a_i}{s+1} \theta_{ij}(s)~ i \ne j\\
	\frac{1}{b_i} \left(\frac{s-a_i}{s+1} \theta_{ii}(s) - \frac{a_i+1}{s+1} \right),~ i = j
	 \end{cases}
	 \end{equation}
which is equivalent to the parameterization \eqref{eq:phiUloc1stOrder}. 
Substituting \eqref{eq:phiu_ij} into \eqref{eq:phix_ij} shows that $\Phi^x$ is of the form \eqref{eq:phiXloc1stOrder}. \hfill $\blacksquare$

\subsection{Proof of Theorem \ref{thm:nthOrderDist}:}

By Lemma \ref{lem:newAffineConstraint}, it is sufficient to show $\Phi^x, \Phi^u \in \oRs$ satisfy \eqref{eq:newAffineConstraint} (for $p = 1$) if and only if they are of the form \eqref{eq:nthOrderDist}. 
The structure of $A$ and $B_2$ allow \eqref{eq:newAffineConstraint} (with $p =1$) to be written by components as:
	\begin{equation}
	(A^{(i)} + I) \Phi^x_{ij} - B^{(i)}_2 \Phi^u_{ij} = \begin{cases} I,~ i = j\\ 0,~ i \ne j \end{cases}
	\end{equation}
For the case $i = j$, parameterizations of $\Phi^x_{ii}$ and $\Phi^u_{ii}$ are then given by Theorem \ref{thm:nthOrder}. A back-stepping approach similar to that used in the proof of Theorem  \ref{thm:nthOrder} can be used to derive a parameterization of $\Phi^x_{ij}$ and $\Phi^u_{ij}$ for the case $i \ne j$. The details of this procedure are omitted.
 \hfill $\blacksquare$

\section{Analytic Solutions for First Order Consensus} \label{app:consensus_computations}

The case of a local error measure with closed-loop band size $M=1$ is analyzed in the following theorem. Similar procedures can be used to derive analogous results for more general choice of $M \ne 1$. 
	\begin{thm} \label{lem:optThetaFirstOrder}
The optimal solution $\theta^{\rm opt} = \{ \theta^{\rm opt}_n(s) \} \in \oRs$ of \eqref{eq:optFirstOrderFinite} with $C = C^{\rm{LE}}$ and band size $M=1$ is given by
	\begin{equation*} \begin{aligned}
	& \theta^{\rm opt}_0(s) =\frac{2 \sqrt{2} - 2 \gamma^2 s + \gamma (s-1)(\alpha + \beta)}{2(\alpha + \gamma s)(\beta + \gamma s) }\\
	& \theta^{\rm opt}_1(s) = \frac{\gamma (\alpha - \beta)(s+1)}{2 \sqrt{2} (\alpha + \gamma s)(\beta + \gamma s)} = \theta^{\rm opt}_{-1}(s) 
	\end{aligned} \end{equation*}
and the resulting optimal closed-loop norm per spatial site is 
	$$
		J~=~\frac{\gamma}{4}(\alpha + \beta), 
	$$
where $ \alpha:=  (2-\sqrt{2})^{1/2} ,~ \beta:=  (2 + \sqrt{2})^{1/2} .$ The closed-loop mappings resulting from $K^{\rm opt}$ are specified by the non-zero entries of their convolution kernels as follows:
	\begin{equation} \begin{aligned} \label{eq:optPhiXPhiU1stOrder}
	&\Phi^x_0 (s)~=~ \frac{4 \sqrt{2} + \gamma (\alpha + \beta)(3s-1) + 2 \gamma^2 s(s-1)}{4(s+1)(\alpha + \gamma s)(\beta + \gamma s)} \\
	&\Phi^x_1(s) ~=~ \frac{\gamma( \alpha - \beta)}{2\sqrt{2} (\alpha + \gamma s)(\beta+ \gamma s)}~ =~ \Phi^x_{-1}(s) \\
	 	& \Phi^u_0(s)~ =~  \frac{\sm (\alpha + \beta) \gamma s (s+3) - 4 \sqrt{2}- 2 \gamma^2 s^2 - 2 \sqrt{2} s}{4(s+1) (\alpha + \gamma s)(\beta + \gamma s )}\\	& \Phi^u_1 (s)~=~ \frac{2 \sqrt{2} + (\alpha + \beta) \gamma s}{2(\alpha + \gamma s)(\beta + \gamma s)}~=~ \Phi^u_{-1}(s).
 	\end{aligned} \end{equation}
	\end{thm}
	\begin{IEEEproof} 
We write \eqref{eq:optFirstOrderFinite} as 
	\begin{subequations}\begin{align} \label{eq:modelMatch1}
	J &=  \inf_{\theta \in \oRs} \small{\Big\|\left[ \begin{array}{c} 
			\frac{1}{s+1} \\ \frac{-1}{s+1} \\ 0 \\0 \\ \hline \frac{-\gamma}{s+1} \\ 0 \\ 0 
			\end{array}\right] 
		+ \frac{1}{s+1} 
		\left[ \begin{array}{ccc} 
				\sm1 & 1 & 0 \\ 
				0 & \sm1 & 1\\ 
				0 & 0 & \sm1 \\ 
				1 & 0 & 0 \\ \hline 
				0 & \gamma s & 0 \\ 
				0 & 0& \gamma s \\ 
				\gamma s & 0 & 0 
			\end{array} \right] 
		\left[ \begin{array}{c}
			\theta_{\sm1} \\ \theta_0 \\ \theta_1 \end{array} \right]   \Big\|_{\h_2}^2} \nonumber \\
			& = : \inf_{\vartheta \in \oRs} ~ \| H + U \vartheta \|_{\h_2}^2\\
			& = \inf_{\vartheta \in \oRs} ~ \| U_i^{\sim} \left( H + U_i U_o \vartheta \right) \|_{\h_2}^2 \nonumber\\ &= \inf_{\vartheta \in \oRs} ~\| U_i^{\sim} H + U_o \vartheta \|_{\h_2}^2 \nonumber \\
			 &=\inf_{m \in \oRs} \| U_i^{\sim} H + m \| _{\h_2}^2 \label{eq:optm}
	\end{align} \end{subequations} 
where we compute an inner-outer factorization $U = U_i U_o$ \cite{francis1987course} as follows. $U_o$ is given as a spectral factor of $U^{\sim} U$:
	\begin{equation*} \begin{aligned}
	U^{\sim} U
		&= \frac{1}{(s+1)(-s+1)}  \left( V \Lambda V^T + -\gamma^2 s^2 I\right)\\
	= ~& \frac{1}{1-s} V (\Lambda^{1/2} - \gamma s I) \cdot \frac{1}{s+1} (\Lambda^{1/2} + \gamma s I)V^*=: U_o ^{\sim} U_o,
	\end{aligned} \end{equation*}
	where $V \Lambda V^*$ is an eigenvector decomposition of $$T:= \left[ \begin{array}{ccc} 2 & \sm1 & 0 \\ \sm1 & 2 & \sm1 \\ 0 & \sm1 & 2 \end{array} \right] = \left[ \begin{array}{ccc} 
				\sm1 & 1 & 0 \\ 
				0 & \sm1 & 1\\ 
				0 & 0 &\sm1 \\ 
				1 & 0 & 0  \ear^T \left[ \begin{array}{ccc} 
				\sm1 & 1 & 0 \\ 
				0 & \sm1 & 1\\ 
				0 & 0 &\sm1 \\ 
				1 & 0 & 0  \ear.$$ $U_i$ is then computed as $U_i:= U U_o^{-1}$. We solve \eqref{eq:optm} with standard projection methods~\cite{luenberger1997optimization}; the optimizer is $$m^{\rm opt} = -(U_i^{\sim}H) \big|_{\oRs},$$ and the corresponding optimal cost is $J= \| \left((U_i^{\sim}H \right)\big|_{\oRs^{\perp}} \|_{\h_2}^2$, where $(\cdot) \big|_{\oRs}$ and $(\cdot)\big|_{\oRs^{\perp}}$ denote projections onto 
$\oRs$ and $\oRs^{\perp}$ respectively. We compute
		\be \begin{aligned} U_i^{\sim}H &=&\lba{c} \frac{\alpha - \gamma}{\sqrt{2} (s+1) } \\ 0 \\ \frac{\beta - \gamma}{\sqrt{2} (s+1) } \ear ~+~& \lba{c} \frac{ \gamma}{\beta (\alpha - \gamma s)} \\ 0 \\\frac{ \gamma}{\alpha (\beta - \gamma s)}\ear \\
		& =&\left( U_i^{\sim}H \right) \big|_{\oRs} ~~+~&~ ~\left( U_i^{\sim}H \right) \big|_{\oRs^{\perp}}
		\end{aligned} \ee
The solution $\vartheta^{\rm opt}$ of \eqref{eq:modelMatch1} is then %
	\begin{equation} \begin{aligned} \label{eq:thetaFromUiUo}
	&\vartheta^{\rm opt} = U_o^{-1} m^{\rm opt} =U_o^{-1} \left( U_i^{\sim}H \right) \big|_{\oRs}\\
		&=\frac{1}{2 (\alpha + \gamma s)(\beta + \gamma s)} \lba{c}\frac{ \gamma }{\sqrt{2}}(\alpha - \beta)(s+1) \\ 2 \sqrt{2} - 2 \gamma^2 s + \gamma (s-1)(\alpha + \beta) \\ \frac{\gamma}{\sqrt{2}}(\alpha - \beta)(s+1) \ear 
	\end{aligned} \end{equation}
	Analytic expressions for the optimal closed-loop maps are determined by $\vartheta^{\rm opt} = \left[\begin{array}{ccc} \theta_{\sm 1}^{\rm opt} & \theta_0^{\rm opt} & \theta_{1}^{\rm opt} \end{array} \right]^T$ using equation \eqref{eq:PhiFromThetaInf} as 
	\begin{equation} \begin{aligned} \label{eq:phiFromSparseTheta}
	\left[ \begin{array}{c} \Phi^x_{-1} \\ \Phi^x_0 \\\Phi^x_{1} \end{array} \right] 
		 &=  \frac{1}{s+1} \left(  \left[ \begin{array}{c} \theta_{-1}\\\ \theta_0  \\\theta_1 \end{array} \right]  + \left[ \begin{array}{c} 0 \\ 1 \\  0 \end{array} \right]
	 \right)\\
	 \left[ \begin{array}{c} \Phi^u_{-1} \\ \Phi^u_0  \\ \Phi^u_{1} \end{array} \right] 
		 &= \frac{1}{s+1} \left(  s\left[ \begin{array}{c} \theta_{-1}  \\ \theta_0  \\ \theta_1 \end{array} \right]  - \left[ \begin{array}{c} 0  \\ 1  \\ 0 \end{array} \right]
	 \right)
	\end{aligned} \end{equation}
to recover equations \eqref{eq:optPhiXPhiU1stOrder}. 
 \end{IEEEproof}

~\\

We next analytically compute the optimal $\theta$ which solves \eqref{eq:optFirstOrderFinite} for varying consensus metrics and band size constraints:

\begin{itemize}
\item \textbf{Local Error, Band Size 2}
\end{itemize}
	\begin{equation*} \begin{aligned}
	 \theta^{\rm opt}_0 &= \frac{\sm1}{3} \Big( \frac{\gamma - r_1 }{r_1 + \gamma s} + \frac{\gamma - \sqrt{2}}{\sqrt{2} + \gamma s} + \frac{\gamma - r_2}{r_2 + \gamma s}\Big)\\
	\theta^{\rm opt}_1 =~ \theta^{\rm opt}_{\sm1} &= \frac{\sm1}{2 \sqrt{3}} \Big( \frac{\gamma -r_1}{r_1 + \gamma s} + \frac{r_2- \gamma }{r_2 + \gamma s} \Big) \\
	\theta^{\rm opt}_2 =~ \theta^{\rm opt}_{\sm2} &= \frac{1}{6} \Big(\frac{r_1- \gamma }{ r_1 + \gamma s } + \frac{2(\gamma - \sqrt{2})}{\sqrt{2} + \gamma s} + \frac{r_2- \gamma }{r_2 + \gamma s} \Big)
	\end{aligned} \end{equation*}
where $r_1:= (2 - \sqrt{3})^{1/2}, r_2:= (2+\sqrt{3}^{1/2})$.
\begin{itemize}
\item \textbf{Deviation from Average, Band Size 1}
\end{itemize}
	\begin{equation*} \begin{aligned}
	\theta^{\rm opt}_0 &= \frac{2(\gamma-1)}{3(1+\gamma s)} - \frac{\sqrt{1 - 3/N} - \gamma}{3 \big( \sqrt{1 - 3/N} + \gamma s \big)}\\
	\theta^{\rm opt}_1=~ \theta^{\rm opt}_{\sm1} &= \frac{\gamma-1}{3 (1 + \gamma s)} - \frac{\sqrt{1 - 3/N} - \gamma}{3 \big( \sqrt{1 - 3/N} + \gamma s \big)}
	\end{aligned} \end{equation*}
\begin{itemize}
\item  \textbf{Deviation from Average, Band Size 2}
\end{itemize}
	\begin{equation*} \begin{aligned}
	\theta^{\rm opt}_0 &=\frac{\sm1}{4} \cdot \frac{1-\gamma}{1 + \gamma s} + \frac{\gamma -\sqrt{1 - 5/N}}{5 \big( \sqrt{1 - 5/N} + \gamma s \big) } \\
	\theta^{\rm opt}_1=~ \theta^{\rm opt}_{\sm1}& = \frac{1}{5}\cdot \frac{1-\gamma}{1 + \gamma s} + \frac{\gamma -\sqrt{1 - 5/N}}{5 \big( \sqrt{1 - 5/N} + \gamma s \big) } \\
	 \theta^{\rm opt}_{2} =~ \theta^{\rm opt}_{\sm2} &= \theta^{\rm opt}_{1}= \theta^{\rm opt}_{\sm1}
	\end{aligned} \end{equation*}
We note that the optimal closed-loop mappings can be recovered from $\theta$ through the formula \eqref{eq:phiFromSparseTheta}.

\section{Consensus: Structured Implementation} \label{app:implementation}

We first form state-space realizations of the 
 transfer matrices made of the concatenation of the 3 non-zero components of $\Phi^x$ and $\Phi^u$ given in Equation \eqref{eq:optPhiXPhiU1stOrder}. A realization of $\lba{ccc} \Phi^x_{\sm 1}(s) & \Phi^x_0(s) & \Phi^x_1(s) \ear$ is:
 		$$
		 \lba{c|c}A_x & B_x \\  \hline C_x & 0 \ear = \lba{ccc} \Phi^x_{\sm 1}(s) & \Phi^x_0(s) & \Phi^x_1(s) \ear,
		 $$
		 \begin{equation*} \begin{aligned}
		A_x &=  \lba{ccc} \sm \frac{\gamma + \alpha + \beta}{\gamma} & 1& 0 \\ \frac{\alpha + \beta}{\gamma}+ \frac{\sqrt{2}}{\gamma^2} & 0 & 1 \\\frac{\sqrt{2}}{\gamma^2} & 0 & 0  \ear , ~ C_x = \lba{ccc} 1 & 0 & 0 \ear,\\
		 B_x &= \lba{ccc} B_{x,1} & B_{x,0} & B_{x,1} \ear \\
		 	& = \lba{ccc} 0 & 2 \gamma^2 & 0 \\
					\gamma (\alpha - \beta) & 3 \gamma (\alpha + \beta) - 2 \gamma^2 & \gamma (\alpha - \beta) \\
					\gamma (\alpha - \beta) & 4 \sqrt{2} - \gamma (\alpha + \beta) & \gamma (\alpha - \beta)
				\ear
	\end{aligned} \end{equation*}
A realization of $ \lba{ccc} \Phi^u_{\sm 1}(s) & \Phi^u_0(s) & \Phi^u_1(s) \ear$ is:
	$$ \lba{c|c}A_x & B_u \\ \hline C_x & 0 \ear = \lba{ccc} \Phi^u_{\sm 1}(s) & \Phi^u_0(s) & \Phi^u_1(s) \ear, $$
where $A_x, C_x$ are as defined above and 
	\begin{equation*} \begin{aligned}
		B_u &= \lba{ccc} B_{u,1} & B_{u,0} & B_{u,1} \ear\\
		& = \lba{ccc} \frac{\alpha + \beta}{2 \gamma} & \sm \frac{\alpha + \beta + 2 \gamma}{4 \gamma } & \frac{\alpha + \beta}{2 \gamma} \\
		\frac{\sqrt{2}}{\gamma^2} + \frac{\alpha + \beta}{2 \gamma} & \sm \frac{3(\alpha + \beta)}{4 \gamma} - \frac{1}{\sqrt{2}\gamma^2} & \frac{\sqrt{2}}{\gamma^2} + \frac{\alpha + \beta}{2 \gamma}\\
		\frac{\sqrt{2}}{\gamma^2} & \sm \frac{\sqrt{2}}{\gamma^2} &\frac{\sqrt{2}}{\gamma^2}
		\ear 
	\end{aligned} \end{equation*}
The local controller dynamics at spatial site $m$ can be written in terms of these matrices as follows:
	\be \begin{aligned}\label{eq:structuredRealization2}
		&\lba{c} \dot{\xi}_m \\ \dot{\zeta}_m \ear = { \lba{cc} A_x + B_{x,0}C_x (A_x + I) & 0 \\ B_{u,0} C_x(A_x + I) & A_x \ear} \lba{c} \xi_m \\ \zeta_m \ear \\
		&~~~+ { \lba{c} B_{x,1} C_x (A_x + I) \\ B_{u,1}C_x (A_x + I)\ear} \left( \xi_{m-1} + \xi_{m+1} \right) +{  \lba{c} B_{x,0} \\ B_{u,0} \ear } x_m \\
		&~~~+ {  \lba{c} B_{x,1} \\ B_{u,1} \ear } (x_{m-1} + x_{m+1})\\
		&u_m = \lba{cc} C_x B_{u,0}  &I \ear C_x (A_x + I)  \lba{c} \xi_m \\ \zeta_m \ear \\&~~~+ C_x B_{u,1}C_x (A_x + I)\left( \xi_{m-1} + \xi_{m+1} \right)  ~+~ C_xB_{u,0} x_m \\
		&~~~+C_x B_{u,1} (x_{m-1} +x_{m+1})
	\end{aligned} \ee
	which is of the form \eqref{eq:structuredRealization} where the state $\psi_m$ of subcontroller $m$ is 
$\psi_m = \left[\begin{array}{cc} \xi_m^T & \zeta_m^T\end{array} \right]^T$, i.e. \eqref{eq:structuredRealization2} is a structured realization with band size $M = 1$.

\ifCLASSOPTIONcaptionsoff
  \newpage
\fi



%\bibliographystyle{IEEEtran}


\begin{IEEEbiography}[{\includegraphics[width=1in,height=1.25in,clip,keepaspectratio]{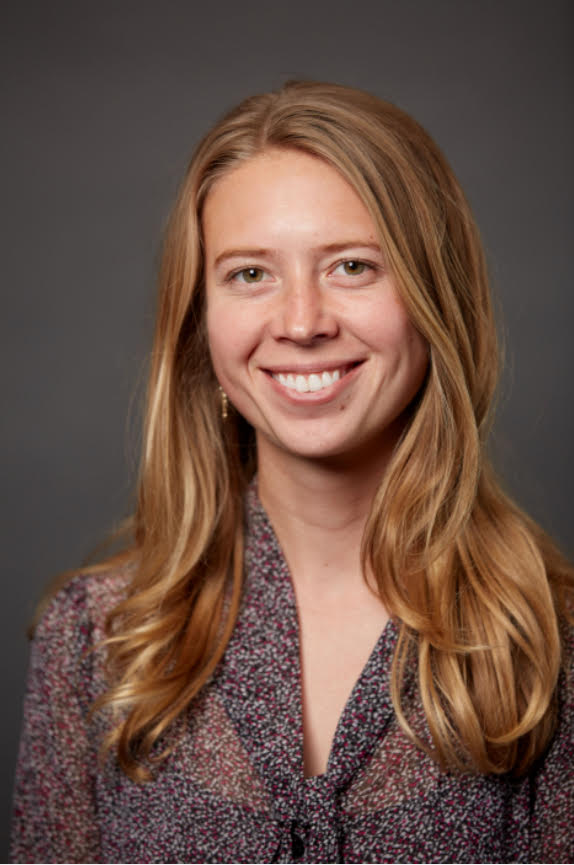}}]{Emily Jensen}
received a B.S. Degree in Engineering Mathematics \& Statistics from the University of California, Berkeley in 2015, after which she worked as a research assistant in the Department of Computing \& Mathematical Sciences at Caltech until beginning her graduate studies in 2016. She received an M.S. and Ph.D. degree in Electrical \& Computer Engineering from the University of California, Santa Barbara (UCSB) in 2019 and 2020, respectively. She is the recipient of the UC Regents' Graduate Fellowship (2016), and of the Zonta Amelia Earhart Fellowship (2019). She is currently a postdoctoral researcher in the Mechanical \& Industrial Engineering department at Northeastern University. 
\end{IEEEbiography}

\begin{IEEEbiography}[{\includegraphics[width=1in,height=1.25in,clip,keepaspectratio]{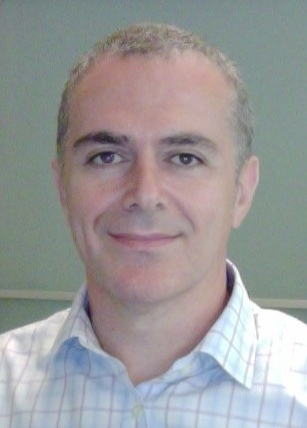}}]{Bassam Bamieh}
(F’08) received the B.Sc. degree in
electrical engineering and physics from Valparaiso
University, Valparaiso, IN, USA, in 1983, and the
M.Sc. and Ph.D. degrees in electrical and computer
engineering from Rice University, Houston, TX,
USA, in 1986 and 1992, respectively. From 1991
to 1998 he was an Assistant Professor with the
Department of Electrical and Computer Engineering,
and the Coordinated Science Laboratory, University
of Illinois at Urbana-Champaign, after which he
joined the University of California at Santa Barbara
(UCSB) where he is currently a Professor of Mechanical Engineering. His research interests include robust and optimal control, distributed and networked
control and dynamical systems, shear flow transition and turbulence, and the
use of feedback in thermoacoustic energy conversion devices. He is a past
recipient of the IEEE Control Systems Society G. S. Axelby Outstanding Paper
Award (twice), the AACC Hugo Schuck Best Paper Award, and the National
Science Foundation CAREER Award. He was elected as a Distinguished
Lecturer of the IEEE Control Systems Society (2005), Fellow of the IEEE
(2008), and a Fellow of the International Federation of Automatic Control
(IFAC).
\end{IEEEbiography}


%\newpage



%\vfill


\end{document}